\documentclass[aps, superscriptaddress, pre, longbibliography]{revtex4-1}
\usepackage{amsmath,graphicx,color,epsfig,latexsym,bm,ulem,dutchcal,subfigure,float,tikz,eucal, mathpazo,times, braket}
\usepackage[colorlinks=true, linkcolor = blue, urlcolor  = blue, citecolor = blue, anchorcolor = red]{hyperref}

\usetikzlibrary{positioning}

\begin{document}
	
	
	\title{Signature of topology via heat transfer analysis in the Su–Schrieffer–Heeger (SSH) model}
	\author{Vipul Upadhyay}
	\affiliation{Department of Physics, Indian Institute of Technology Delhi, Hauz Khas 110 016, New Delhi, INDIA,}
	\author{M. Tahir Naseem} 
	\affiliation{Faculty of Engineering Science, Ghulam Ishaq Khan Institute of Engineering Sciences and Technology, \\
		Topi 23640, Khyber Pakhtunkhwa, Pakistan,}
	\author{\"Ozg\"ur E. M\"ustecapl\ifmmode \imath \else \i \fi{}o\ifmmode \breve{g}\else \u{g}\fi{}lu}
	\email{omustecap@ku.edu.tr}
	\affiliation{Department of Physics, Ko\c{c} University, 34450 Sariyer, Istanbul, T\"urkiye,}
	\affiliation{T\"UB\.ITAK Research Institute for Fundamental Sciences, 41470 Gebze, T\"urkiye.}
	\affiliation{Faculty of Engineering and Natural Sciences, Sabanc\i ~University, Tuzla 34956, Istanbul, T\"urkiye.}
	\author{Rahul Marathe}\email{maratherahul@physics.iitd.ac.in} \affiliation{Department of Physics, Indian Institute of Technology Delhi, Hauz Khas 110 016, New Delhi, INDIA,}
	
	\begin{abstract}{ {In this work, we explore how thermodynamics can be a potential tool for identifying the topological phase transition}. Specifically, we focus on a one-dimensional Su–Schrieffer–Heeger (SSH) chain sandwiched between two fermionic baths. To investigate distinctive thermodynamic signatures associated with the topological phase, we employ heat flow analysis. Our results, derived using a global master equation, unveil a significant suppression of heat flow as we transition from the trivial to the topological phase. This decline in heat flow can be attributed to the reduction in transmission coefficients of non-zero energy modes within the topological phase. It may serve as an indicator of a phase transition. Furthermore, we investigate the heat flow asymmetry to search for phase transition indicators. Interestingly, no asymmetry is observed when employing fermionic baths. However, upon substituting fermionic baths with bosonic ones, we report a non-zero heat flow asymmetry. For SSH model with few fermionic sites, this asymmetry is more pronounced in the topological phase compared to the trivial phase. Therefore, the observed behavior of the heat diode provides an additional means of distinguishing between the topological and trivial phases. Finally, we delve into the contributions from both bulk and edge effects in heat flow and rectification to explore the impact of small system sizes on our findings.}
	\end{abstract}
	
	\maketitle

	\section{Introduction}
	
	In classical physics, phase transitions are identified through the examination of thermodynamic signatures, manifesting as abrupt changes in properties like specific heat or magnetization. However, quantum phase transitions, distinct from classical ones, occur solely due to quantum mechanical properties, particularly at absolute zero temperatures. These transitions give rise to exotic quantum phases, emphasizing the importance of identifying and understanding them. This unique realm of condensed matter physics has become a focal point of research \cite{PhysRevB.74.174422, PhysRevLett.115.177204, PhysRevLett.118.267701, PhysRevB.99.064422, PhysRevE.102.032127, PhysRevB.105.L060401}, demanding innovative approaches to detect and comprehend these transitions, as they hold the key to unlocking novel quantum materials and technological advancements.

	Transport studies in topological materials have been a subject of intense research \cite{Sensing_Floquet, Charge_and_spin_transport,driven_topology,edge_state_entropy_Mondal_2023,signature_dynamic_driven_kitaev}. This is because the separation of edge and bulk in the topological phase leads to exotic transport properties for a system in topological phase \cite{nature_main_ref,Sensing_Floquet}. In mathematics, topology is a geometrical property for any system \cite{Kane_topo_review} which is robust against continuous deformation. Depending upon the context of the specific system, a topological invariant may be mathematically defined. In quantum mechanics, the topology appears due to the independence in defining the phase of the wavefunction. As such we can define the topology of energy band structures in a quantum system. It is a property that is used to distinguish between two phases of a Hamiltonian which can't be smoothly transformed into each other without crossing the energy gap between ground and bulk. These are also called non-adiabatically connected regions \cite{Kane_topo_review, short_course_topo}. For isolated systems at absolute zero, winding number or Berry phase \cite{short_course_topo}  plays the role of topological invariant helping us to distinguish between the phases of the system. Coupling the system to heat baths will usually not preserve the topological invariants and new topological invariants like the Ulham phase are needed \cite{Ulham_ref,hill_kempkes_ref}. In 1-D quantum systems topological phase transition results in localization of wavefunction at edges. This localization can then be utilized for altering transport properties and indicators of the phase of the system.  A lot of studies have been dedicated to establishing heat transport as an indicator of the topological phase. Experiments for such systems are possible in cold atoms, and nanowires  \cite{exp1, exp2, exp3, exp4, exp5}.

	In this study, we set out to address two main inquiries: firstly, the viability of utilizing heat transport for identifying the topological phase within the SSH model; and secondly, the prospect of harnessing the asymmetry in heat transport across different phases to engineer a specialized device, notably a heat diode \cite{topo_diode1,topo_diode2,topo_diode3,topo_diode_liquid}, whose operational efficiency hinges upon the specific topological phase encountered. Our investigation yields affirmative responses to both questions, subject to defined parameter regimes. Additionally, we analyze thermodynamic intricacies associated with topological phase transitions. This analysis entails the identification of an effective temperature corresponding to a specific energy mode of the system, following the analytical framework proposed by previous studies \cite{hill_kempkes_ref,Quelle_PhysRevB.94.075133}. Notably, our analysis differs from prior works by adopting a global master equation, thus ensuring thermodynamic consistency and obviating the reliance on the local master equation \cite{similar_2,similar_ssh} which may violate the second law of thermodynamics \cite{Levy_2014,PhysRevA.98.052123}. Furthermore, our method does not require external chemical potentials at each site, distinguishing it from previous studies. For the rectification analysis \cite{diode1,diode2,diode4,diode6,dvira_segal_diode}, we replace fermionic baths with bosonic counterparts and exploit inherent bath-system interaction asymmetry to induce rectification. While this technique echoes prior studies in inducing rectification \cite{dvira_segal_diode}, its application as a potent harbinger of phase transitions within a one-dimensional SSH chain represents uncharted territory within our study.

	\par The structure of this paper is as follows: In Sec.~\ref{Sec: SSh_sec}, we present the model and introduce the global master equation. Sec.~\ref{subsec:Results} is dedicated to the analysis of heat flow results. In Sec.~\ref{Sec:Effective_temp}, we define the effective temperature and explore the thermodynamic characteristics of our system. We replace the fermionic baths with bosonic baths and investigate heat current rectification in Sec.~\ref{Sec:diode}. To conclude, in Sec.~\ref{sec:conclusion}, we summarize our findings and offer concluding remarks. In the appendices \ref{section:Appendix A} and \ref{Appendix:B} we give detailed derivations of some of the results used in the main text. In Appendix \ref{Appendix:C} we provide some additional supporting results.
	
	\section{The SSH Model} \label{Sec: SSh_sec}
	
	\begin{figure}[!t]
		\begin{center}
			\includegraphics[width=0.5\textwidth,angle=0]{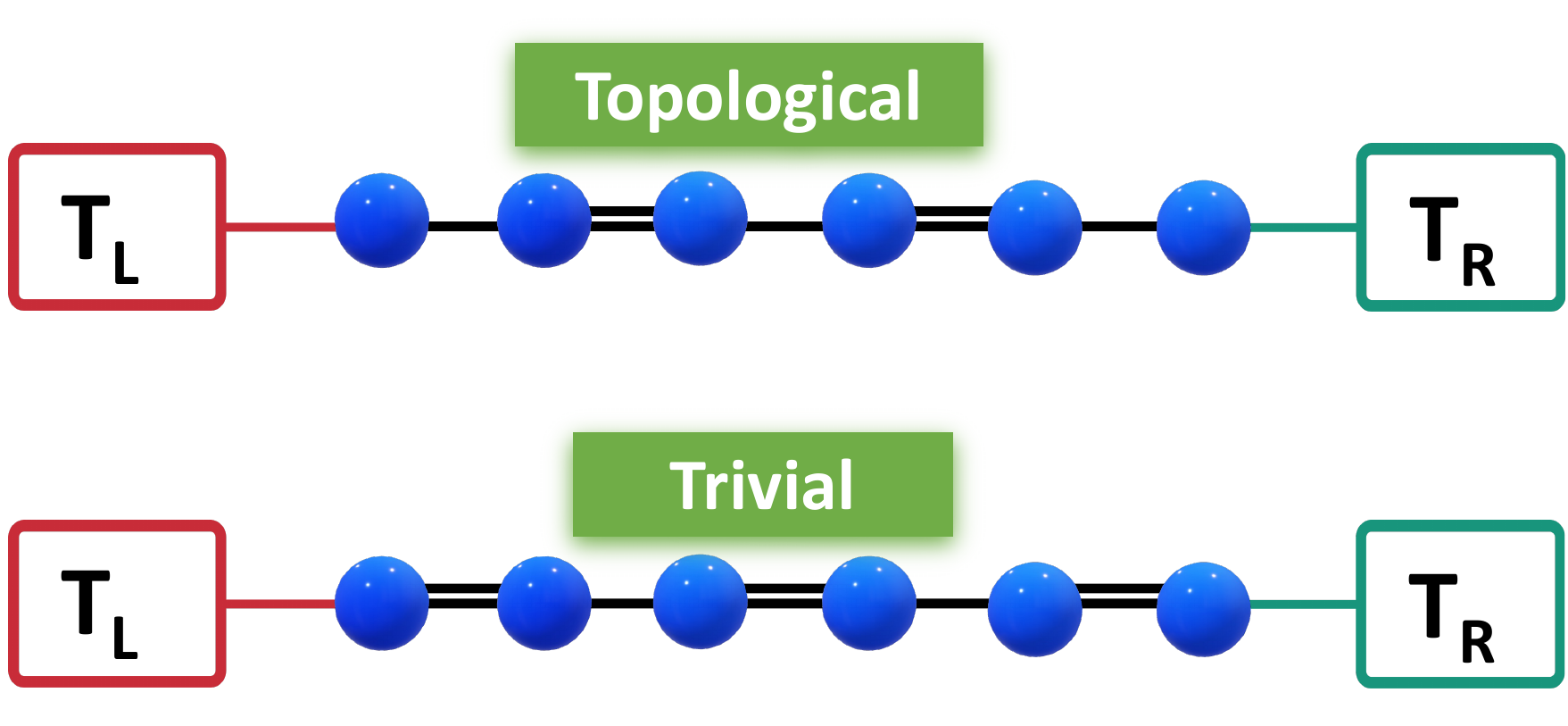}
			\caption{{Schematic of the model system. A single SSH chain is connected at left and right ends to two heat baths at temperature $T_L$ and $T_R$ respectively. Double and single bonds signify the relative hopping strengths  between sites with double bonds having larger hopping strength than the single bond.}}
		\end{center}
	\end{figure}
	In our model, we consider a one-dimensional system consisting of fermions occupying $N$ sites. The interaction strengths of these fermions vary depending on whether the bonds between sites are odd or even. It is important to note that $N$ is always chosen to be an even number. The Hamiltonian governing this model is expressed as follows \cite{ssh_original_paper, Ssh_eigen}:
	\begin{align}\label{model_hamiltonian}
		\hat{H}_S&=-t\sum_{n}^{N-1} (1+(-1)^n \delta) (\hat{c}_{n+1}^\dagger \hat{c_n}+\hat{c}_{n}^\dagger \hat{c}_{n+1})
	\end{align}
	where,
	$v_1=-t(1-\delta)$, $v_2=-t(1+\delta)$	are the odd and even bond strengths, respectively. This represents the simplest 1-D model 
	permitting a topological phase. It is well known \cite{ssh_original_paper,Ssh_eigen} that when $v_1<v_2$, the system enters the topologically non-trivial phase, whereas the phase characterized by $v_2<v_1$ is the topologically trivial phase. For convenience, we refer to the two phases as topological phase and trivial phase respectively. Consequently, $\delta$ can be interpreted as a parameter indicative of topological order, wherein $\delta < 0$ signifies the trivial phase, while $\delta > 0$ corresponds to the topological phase. The primary distinction between the trivial and the topological phase is the 
	existence of two zero-energy modes within the latter \cite{short_course_topo}. This difference leads to the localization of energy eigenvectors corresponding to these zero energy modes at the two ends of the chain. 
	This system is coupled at its two ends with two distinct fermionic baths that do not overlap. Each of these baths is governed by its respective Hamiltonian
	\begin{align}
		\hat{H}_B=\sum_{n}\omega^L_n \hat{b}_{n}^{\dagger L}\hat{b}^L_{n}+\sum_{n}\omega^R_n \hat{b}^{\dagger R}_{n } \hat{b}^R_{n},
	\end{align}	
	where {$\hat{b}_{n}^{\dagger i} (\hat{b}^i_{n})$ are the creation (destruction) operator for the $n$-th mode of the $i$-th ($i = L, R$) bath  and $w^i_n$ is the energy corresponding to the $n$-th mode  of the $i$-th bath. The system bath interaction is  of the full coupling form \citep{System_Bath_PhysRevA.102.022209,System_bath_SARGSYAN2018666,System_bath_PhysRevE.97.032134}. This interaction is chosen because it is a more general form of the system-bath interaction. The system-bath Hamiltonian then reads}
	\begin{align}
		\hat{H}_{SB}&=(\hat{c}_1^{\dagger}+\hat{c}_1)\otimes\sum_k (g^L_k \hat{b}^L_k+g^{ *L}_k\hat{b}_k^{\dagger L})+(\hat{c}_N^{\dagger}+\hat{c}_N)\otimes \sum_k (g^R_k \hat{b}^R_k+g^{ *R}_k\hat{b}_k^{\dagger R}),
	\end{align}
	where {$g^L_k$  and $g^R_k$ are interaction strengths of the system with the $k$-{th} mode of the left and right bath respectively. We work in the units where $k_B=1,\hbar=1$}.
	\subsection{System eigenvalues and eigenvectors} \label{eigesnsystem}
	The eigenvalues of the SSH model are presented as \cite{Ssh_eigen,ssh_eigen_2}
	\begin{align}
		\lambda_k=\pm \sqrt{2}t\sqrt{1+\delta^2+(1-\delta^2)\cos{\theta_k}}
	\end{align}
	where $\theta_k$ can be found by solving the  equation
	\begin{align}
		(1-\delta) T(\theta_k,N/2)+(1+\delta)T(\theta_k,N/2-1)=0,
	\end{align}
	where,
	\begin{align}
		T(\theta_k,n)=\frac{\sin{[(n+1)\theta_k]}}{\sin{\theta_k}}.
	\end{align}
	The equation above has $N$ real solutions when $\delta < 0$. However, with $\delta > 0$, its behavior changes: it includes ($N - 2$) real solutions along with 2 complex roots. These complex roots correspond to the zero energy modes of the SSH model, which are linked to the complex solutions of $\theta_k$.	
	The elements of the eigenvector  $\boldsymbol{\Phi}^k$ associated with the  eigenvalue $\lambda_k$  are given as \cite{Ssh_eigen}:
	\begin{align} \label{eigenVector}
		\boldsymbol{\Phi}_{2n-1}^k&= \frac{1+\delta}{1-\delta} T(\theta_k,n-2)+T(\theta_k,n-1),\nonumber
		\\ \boldsymbol{\Phi}_{2n}^k&=\pm \frac{\lambda_k}{t(1-\delta)} T(\theta_k,n-1).
	\end{align}
	Considering that $t$ serves as a mere multiplication factor in our Hamiltonian, the eigenvectors should remain independent of the value of $t$. Moreover, owing to the symmetries inherent in the Hamiltonian, we can expect additional simplifications in our eigenvectors \cite{similar_ssh}. 
	As the Hamiltonian in Eq. \eqref{model_hamiltonian} exhibits symmetry under the transformation  $n\to N+1-n$, we find that  
	$|\boldsymbol{\Phi}^k_{1}|^2=|\boldsymbol{\Phi}^k_{N}|^2$ for non near-zero energy modes. For zero-energy modes, wavefunction localization occurs at either end, rendering this symmetry inapplicable.
	Finally, as a result of particle-hole symmetry, we have $|\boldsymbol{\Phi}^{k}_{n}|^2=|\boldsymbol{\Phi}^{-k}_{n}|^2$.

	
	\subsection{Master equation and heat flow}
	Utilizing the eigenvectors mentioned above and constructing a unitary matrix $U$, we can diagonalize the system's Hamiltonian and express it in the energy basis as (see Appendix \ref{section:Appendix A} for details)
	\begin{align}
		\hat{H}_S=\sum_k \omega_k\hat{a}^{\dagger}_k \hat{a}_k,
	\end{align}
	where, $\omega_k$ represents the energy associated with the $k$-th mode, while $\hat{a}^{\dagger}_k$ and $\hat{a}_k$ denote the corresponding creation and destruction operators, respectively.
	The derivation of the master equation for our model adheres to the conventional Born-Markov and secular approximations. In the Schr\"{o}dinger picture, it is given by~\cite{breuer2002}
	\begin{equation}\label{liouville}
		\frac{d}{dt}\hat{\rho}=-i[\hat{H}_S,\hat{\rho}]+\mathcal{L}_{L}\hat{\rho}+\mathcal{L}_{R}\hat{\rho},
	\end{equation}
	where, $\hat{\rho}$ is the density matrix of the N site SSH model, and the Liouvillian super-operators of the left and right baths are (see Appendix \ref{section:Appendix A}),
	\begin{align}\label{left_dissipator}
		\mathcal{L}_{L}\hat{\rho}&=\kappa^L_j \sum_{j} G(\omega_j,T_L) \mathcal{D}(\hat{a}_j)[\hat{\rho}]+ G(-\omega_j,T_L)\mathcal{D}(\hat{a}_j^{\dagger})[\hat{\rho}],\nonumber \\
		\mathcal{L}_{R}\hat{\rho}&= \kappa^R_j \sum_{j}G(\omega_j,T_R) \mathcal{D}(\hat{a}_j)[\hat{\rho}]+ G(-\omega_j,T_R)\mathcal{D}(\hat{a}_j^{\dagger})[\hat{\rho}],
	\end{align}
	The index $j$ in the sum runs over all the eigenvalues,
	$\kappa^L_j=\kappa_L |U_{1,j}|^2$, and $\kappa^R_j=\kappa_R |U_{N,j}|^2 $ and { $\kappa_{L(R)}$ is proportional to the square of system bath interaction strength ($\kappa_{L(R)}\propto |g^{L(R)}_k|^2$). The complete characterization of $\kappa_{L(R)}$ depends on the choice of the spectral densities of the bath \cite{breuer2002}. The details are provided in the appendix \ref{section:Appendix A}, Eq. \eqref{J_definition}.}
	\par Considering that $|U_{i,j}|^2$ represents the square of the $i$-th component of the $j$-th eigenvector, we can express $\kappa^L_j$ $(\kappa^R_j)$ in terms of the eigenvectors defined in Eq. \eqref{eigenVector}. Specifically, $\kappa^L_j=\kappa_L |\boldsymbol{\Phi}^j_{1 (N)}|^2$ and $\kappa^R_j=\kappa_R |\boldsymbol{\Phi}^j_{1(N)}|^2 $. Here, we have leveraged the symmetry property as outlined in Sec. \ref{eigesnsystem}. The Lindblad dissipators given in Eq. \eqref{left_dissipator} are defined as
	
	\begin{align}
		\mathcal{D}(\hat{A})[\hat{\rho}]=2\hat{A}\hat{\rho}\hat{A}^\dagger-\hat{A}^\dagger\hat{A}\hat{\rho}-\hat{\rho}\hat{A}^\dagger\hat{A},
	\end{align}
	and bath spectral density is given by 
	\begin{align}
		G(\omega,T_i)=
		\begin{cases}
			J(\omega)(1-{f}_i(\omega) )& \omega>0, \\
			J(|\omega|){f}_i(|\omega|) & \omega<0.
		\end{cases}
	\end{align}
	{ In this work, for convenience, we assume flat spectra for the baths, implying that $J(\omega)= 1$ holds true for all $\omega$ values. The particular choice of the bath spectrum function does not alter the qualitative behavior of the heat current as illustrated in the Appendix \ref{Appendix:C}.}
	The Fermi-Dirac distribution is given as
	\begin{align}\label{eq:fermi-dirac}
		{f}_i(\omega)=\frac{1}{e^{\omega/T_i}+1},
	\end{align}
	{ where we have assumed equal chemical potentials for both baths, setting them to `0' for the sake of convenience.}
	Finally, the heat current from the left and right baths can be calculated using \cite{nature_main_ref}
	\begin{align}\label{eq:heat}
		I_{L (R)}=\text{Tr}(\mathcal{L}_{L (R)} H_S).
	\end{align}
	
	\section{Results}\label{subsec:Results}

	\begin{figure}[t]
		\centering 
		\subfigure []
		{\includegraphics[width=0.32\linewidth,height=0.25\linewidth]{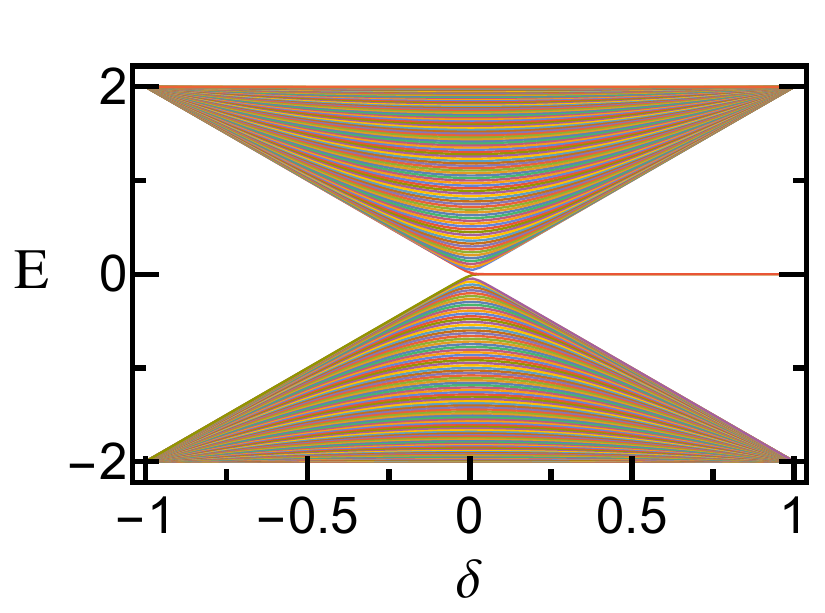}}
		\subfigure []
		{\includegraphics[width=0.32\linewidth,height=0.25\linewidth]{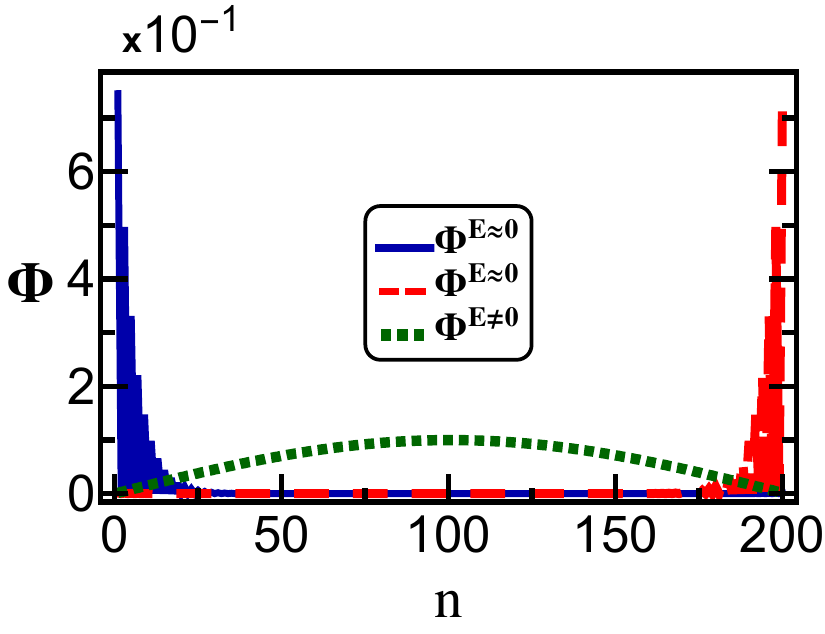}}
		\subfigure []
		{\includegraphics[width=0.32\linewidth,height=0.25\linewidth]{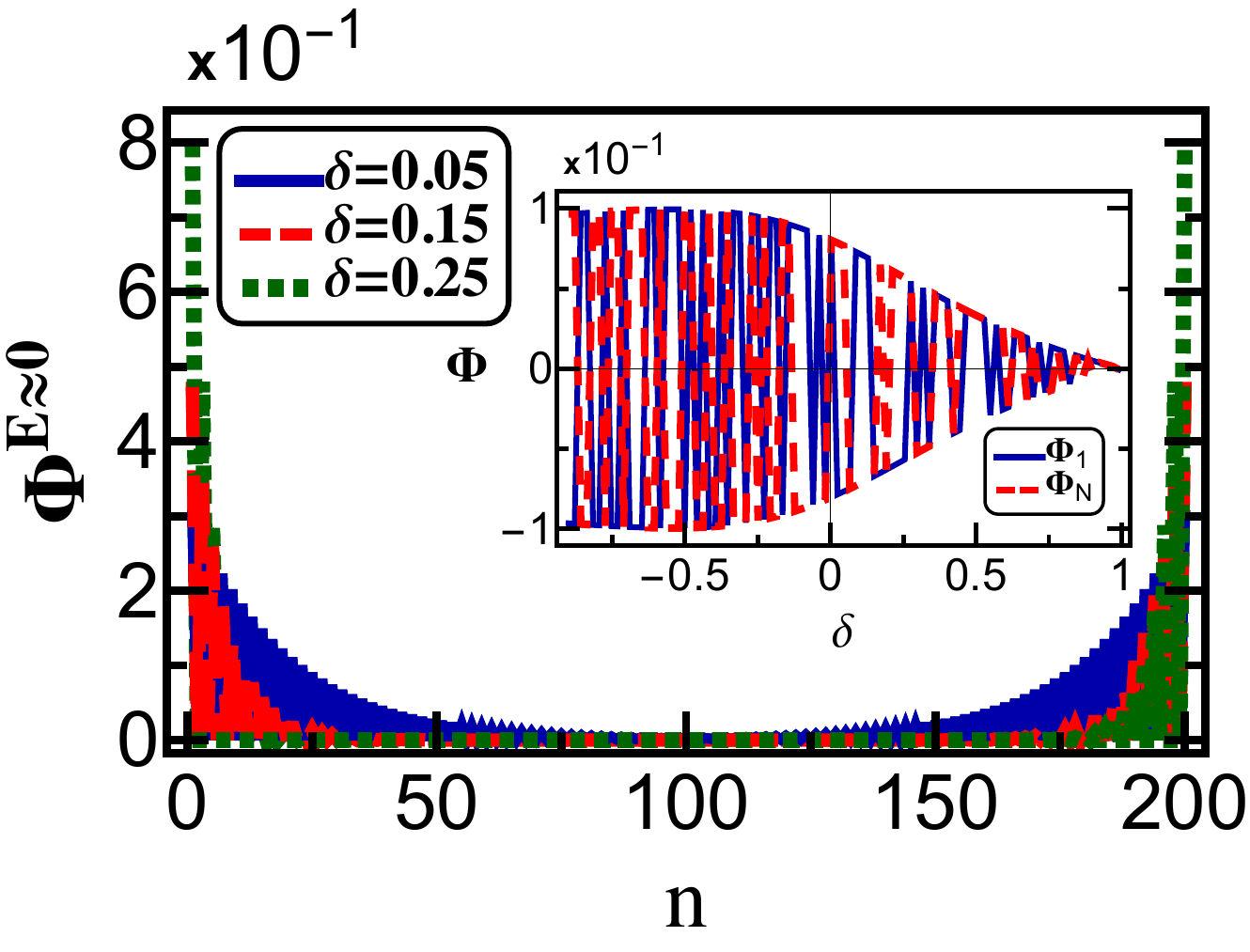}}
		\subfigure []
		{\includegraphics[width=0.32\linewidth,height=0.25\linewidth]{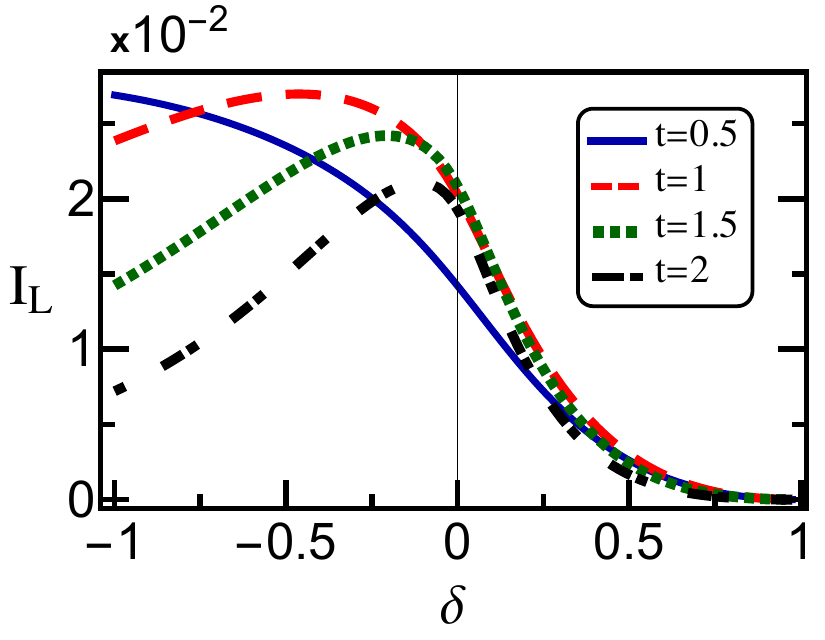}}
		\subfigure[]
		{\includegraphics[width=0.32\linewidth,height=0.25\linewidth]{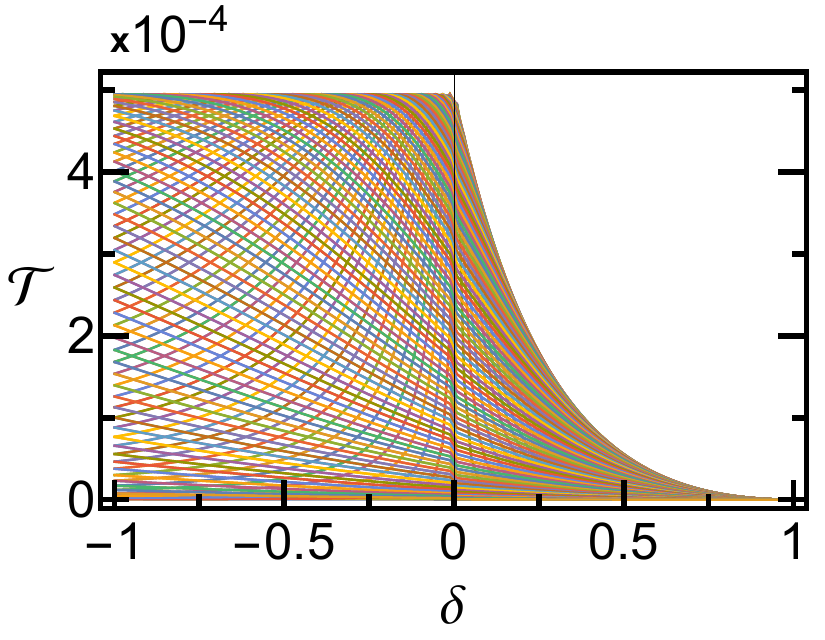}}
		\subfigure[]
		{\includegraphics[width=0.32\linewidth,height=0.25\linewidth]{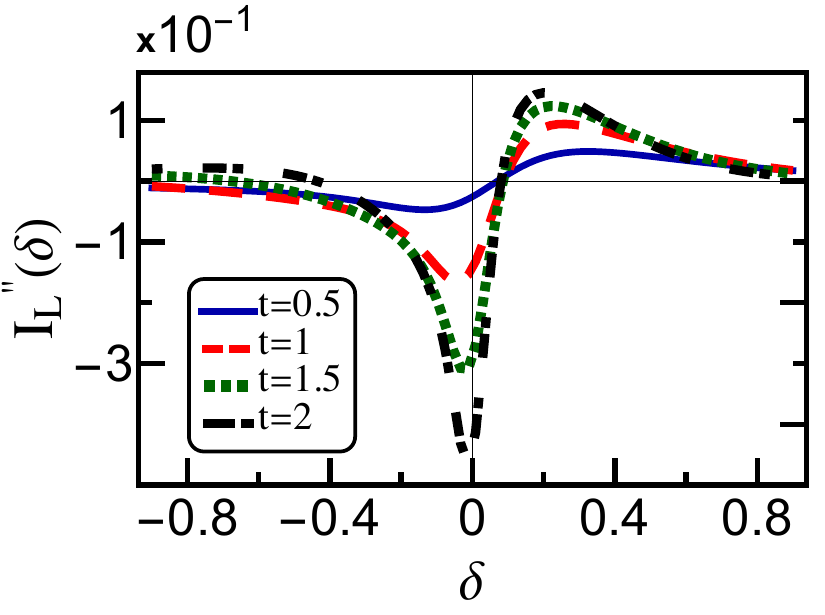}}
		\caption{{\textbf{(a)} Energy spectrum with $\delta$.  Variation of absolute value of wavefunction at site $n$ for \textbf{(b)} near zero as well as non-zero energy states, \textbf{(c)} near zero energy states for various $\delta$, (inset) variation of first and last component of eigenvector of a typical non-zero energy mode with $\delta$,	\textbf{(d)} Variation of heat current with $\delta$ for various $t$'s, \textbf{(e)} variation of transmission coefficients for all the non-zero energy modes with $\delta$ and \textbf{(f)} variation of the curvature of the heat current with $\delta$. For all the case, if not specified otherwise, the values of the parameters are $N=200, t=1, T_L=1, T_R=0.1, \delta=0.2, \kappa_L=0.1, \kappa_R=0.1$.}}
		\label{current_delta_figures}
	\end{figure}
	{
		We begin our analysis by revisiting the topological features of the SSH model. In Fig. \ref{current_delta_figures}(a), we plot the energy spectrum of the SSH model as a function of the order parameter, $\delta$. We observe that a substantial portion of the energy eigenvalues lies above the $`2 \delta'$ threshold in both the phases and the primary spectral distinction between the trivial and topological phases is the presence of two nearly zero-energy modes in the topological phase. Plotting the wavefunction for the two near zero energy modes in Fig. \ref{current_delta_figures} (b), we see that these two wavefunctions localize at the first and last sites in the topological phase. As a result of this localisation, the components of non-zero energy modes in the first and last site are very small in the topological phase. In Fig.  \ref{current_delta_figures} (c), we see that the extent of localisation of the near zero energy modes increases as we increase the value of $\delta$ and go deeper into the topological phase. This is expected because in the extreme limit of $\delta\to1$, it can be shown \cite{short_course_topo} that the fermionic sites at the edges separate from the bulk of the chain and these sites  support zero energy modes. Correspondingly, this also means that the component of non-zero energy states on the first and last sites continuously decrease on going deeper inside the topological phase as can be seen in the inset of the Fig. \ref{current_delta_figures} (c). Examining all these results, we suspect that the localisation of the zero energy modes at the edges can hinder the transfer of heat current between the baths in the topological phase. This can happen because the wavefunctions at the edges are not sufficiently spread out to transfer the heat from the baths to the bulk of the SSH chain. To gain deeper insights into this,  we perform a thorough heat transfer analysis below. 
		\subsection{Fall in heat current in the topological phase:}
		The expression for the heat current, as computed using Eq. \eqref{eq:heat}  for the SSH model, is given by (see Appendix \ref{Appendix:B})
		\begin{align} \label{heat_expression}
			I_L=4\sum_{\omega_j>0}\mathcal{T}_j\omega_j ({f}_L(\omega_j)-{f}_R(\omega_j)).	 
		\end{align}
		We observe that this expression for heat current adheres to the Landauer form \cite{dvira_segal_diode}. This representation unveils that the total heat current arising from the summation of heat currents flowing through independent energy channels with frequencies $\pm \omega_j$. Additionally, each energy channel $\omega_j$ is characterized by its respective transmission coefficient, denoted as $\mathcal{T}_j$ \cite{datta_quantum_transport}
		\begin{align}\label{transmission_coefficients}
			\mathcal{T}_j= \frac{k^L_jk^R_j}{k^L_j+k^R_j}=|\boldsymbol{\Phi}^j_{1 (N)}|^2 \frac{\kappa_L \kappa_R}{\kappa_L+\kappa_R}.
		\end{align}
		Looking at the variation of heat current with $\delta$ in Fig. \ref{current_delta_figures}(d),  it becomes apparent that the heat current in the topological phase experiences suppression compared to the trivial phase.
		\\ 
		Examining the heat current expression in Eq. \eqref{heat_expression}, we observe two distinct factors influencing its variation at fixed bath temperatures. The first factor pertains to alterations in energy levels, while the second factor relates to modifications in transmission coefficients. Upon closer examination of Fig. \ref{current_delta_figures}(a), it becomes evident that all heat conduction channels on either side are the same, with the exception of two near-zero energy channels within the topological phase. However, it's worth noting that these near-zero energy channels make only negligible contributions to heat current in large systems. Consequently, this difference in energy spectrum diminishes as we scale up the system size. Therefore, changes in energy levels alone can not account for the substantial decrease in heat current observed in the topological region.
		Now we analyze the transmission coefficients associated with energy levels greater than $2\delta$, as defined in Eq. $\eqref{transmission_coefficients}$. As we go deeper into the topological phase, the wavefunctions of the two zero-energy modes gradually localize toward the two ends of the chain. Consequently, the first and $N$-th components of non-zero eigenvectors begin approaching zero. The further we advance into the topological region, the more pronounced this effect becomes { because the localisation of near zero energy modes increases at the edges}. Consequently, the transmission coefficients for all bulk energy channels gradually tend toward zero, resulting in a significant reduction in heat current within the topological region. This observation aligns with previous findings reported in \cite{similar_ssh}. 
		
		In Fig. \ref{current_delta_figures}(e), we present the bulk transmission coefficients for all available channels. In the trivial phase, the average of all these transmission coefficients remains almost the same, but as we transition into the topological phase, it starts to decrease noticeably. This tendency of near-zero energy modes to localize at the ends is an inherent feature of the topological property characterizing the SSH model. Therefore, we can confidently assert that the reduction in heat current observed in the topological region is a direct consequence of the system's topological characteristics.
		Though, the heat current does decrease as we enter the topological phase, its behavior alone can not precisely indicate the point of phase transition as the decline in heat current is rather smooth. Therefore,  the heat current behavior  should be studied in conjunction with the behavior of transmission coefficients. On careful observation of the behavior of heat current near $\delta=0$,  motivated us to look at the curvature of the heat current as a function of $\delta$. To check whether the curvature could be a better indicator of the phase transition, we plot the second derivative of the heat current with $\delta$ in Fig. \ref{current_delta_figures}(f) and see that it is indeed a much better indicator of phase transition with the minima of the curvature almost coinciding with the phase transition point. We also see that as we increase the intrasystem interaction strength $t$, the ability of the curvature to indicate phase transition becomes more and more accurate. This may be because the global master equation used by us in this manuscript gives more reliable results for higher intrasystem coupling strengths.    Notably, the current does not approach zero as the bond strength significantly weakens in the trivial phase; instead, it saturates at $`2 t (f_L(2t)-f_R(2t))'$. This is attributed to the breakdown of the secular approximation in the derivation of the global master equation, particularly when dealing with a weak intrasystem coupling strength  \cite{global_issue}. Therefore, we limit our analysis to scenarios in which $\kappa <<t(1\pm\delta)$ to ensure the continued validity of the secular approximation. As already mentioned, these results are obtained for flat spectrum of the bath. However, the particular choice of the bath spectrum function does not alter the qualitative behavior of the heat current as illustrated in the Appendix \ref{Appendix:C}.}
	\begin{figure}[t]
		\centering 
		\subfigure []
		
		{\includegraphics[width=0.35\linewidth,height=0.26\linewidth]{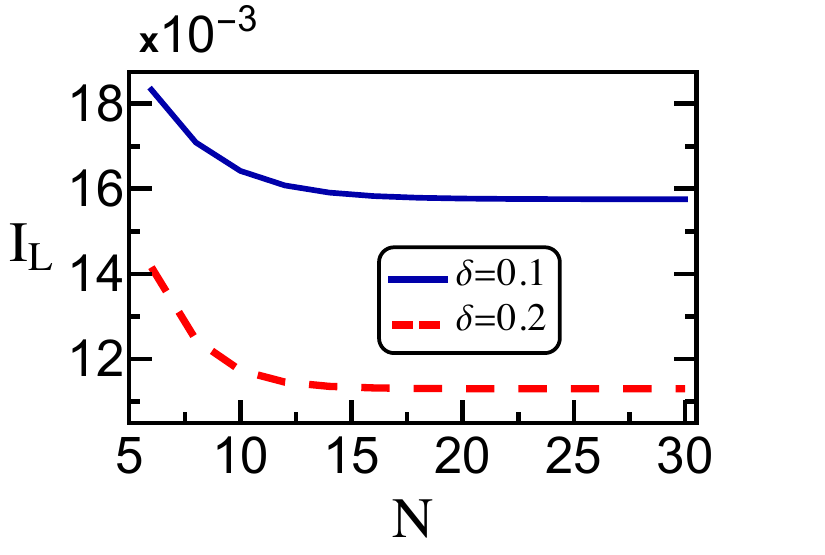}}
		\subfigure []
		{\includegraphics[width=0.31\linewidth,height=0.25\linewidth]{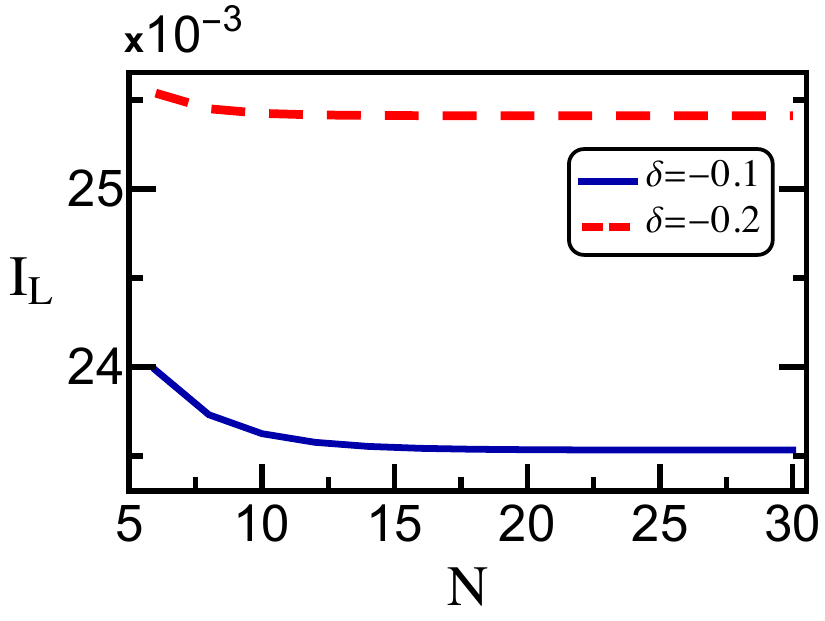}}
		\subfigure []
		{\includegraphics[width=0.32\linewidth,height=0.25\linewidth]{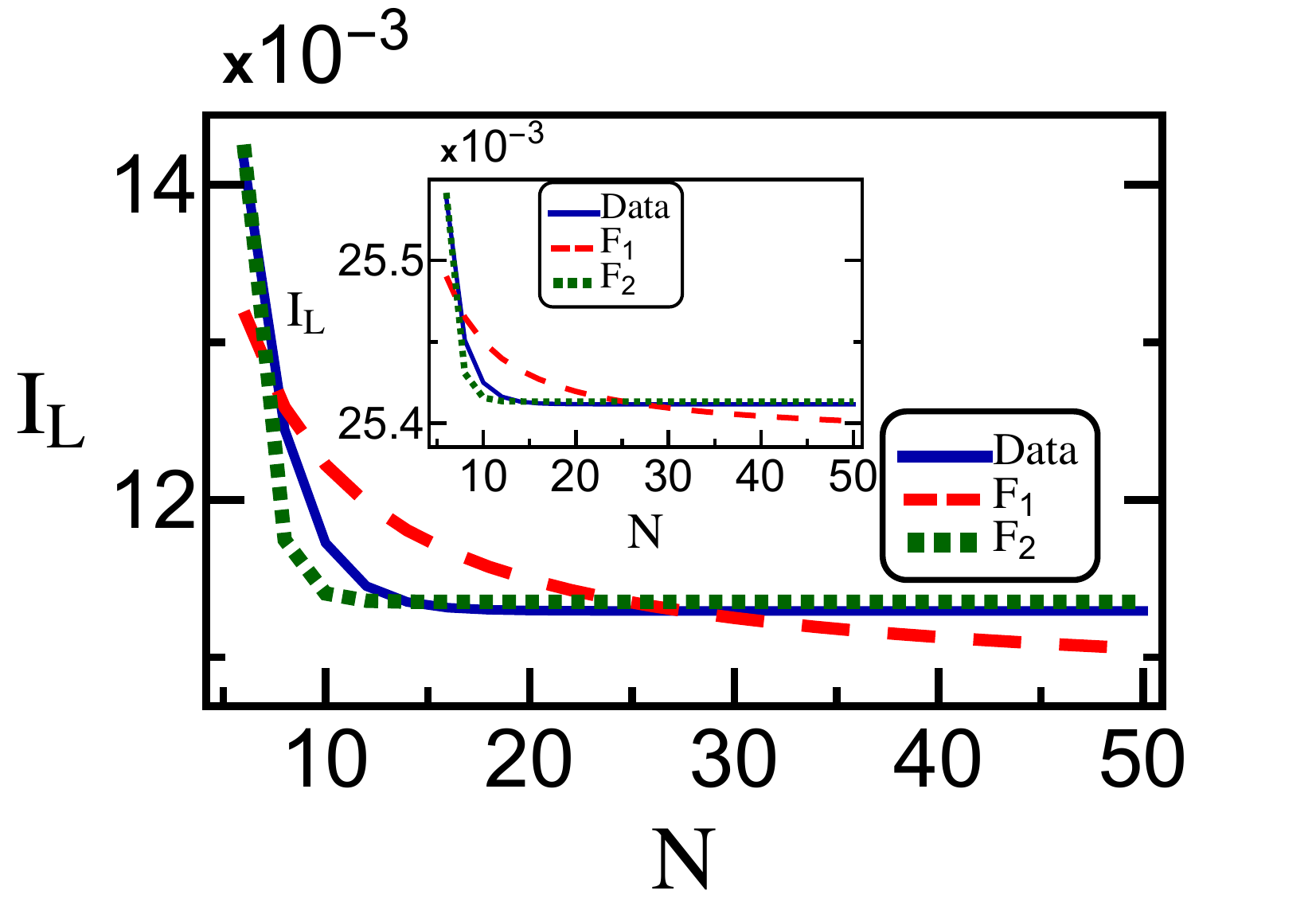}}
		\subfigure []
		{\includegraphics[width=0.32\linewidth,height=0.25\linewidth]{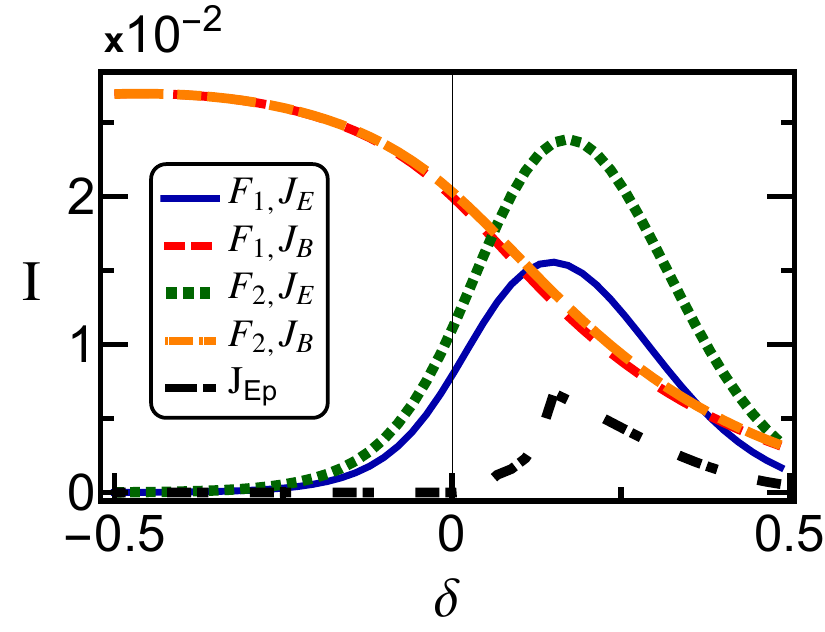}}
		\subfigure []
		{\includegraphics[width=0.38\linewidth,height=0.26\linewidth]{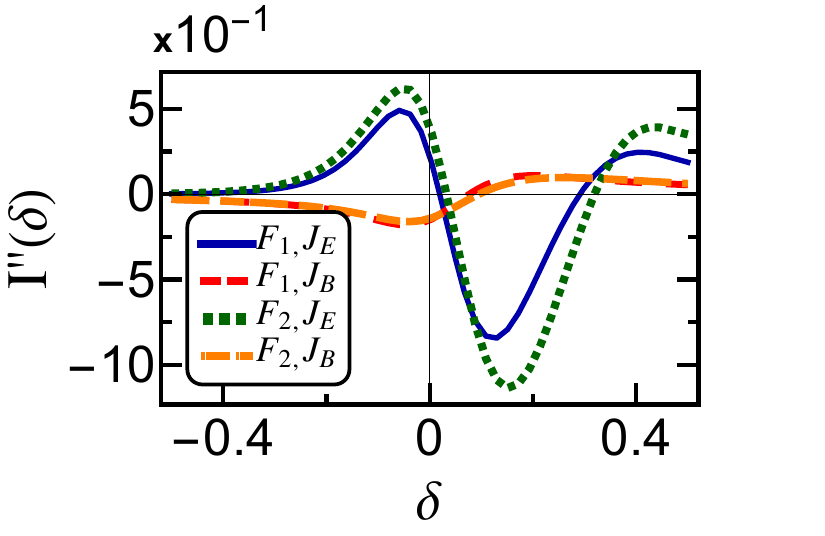}}
		\caption{Variation of heat current with $N$ for \textbf{(a)} Topological Phase, \textbf{(b)} Trivial phase. \textbf{(c)} Comparison of fitting quality of $F_1$ and $F_2$ for variation of heat current with $N$ for topological and trivial (inset) phase. \textbf{(d)} Variation of edge and bulk current with $\delta$.
			The edge current in fit $F_2$ has been scaled by a factor of `$\frac{1}{50}$' for visual convenience. All the fits are done for data in the range $N=\{6,50\}$. 
			{\textbf{(e)} Variation of curvature of edge and bulk current with delta.}}
		\label{Edge_and_bulk_fig}
	\end{figure}
	
	\subsection{Heat current with system size}
	
	Examining the behavior of heat current concerning the system size N in Figs. \ref{Edge_and_bulk_fig}(a) and \ref{Edge_and_bulk_fig}(b). When $N$ exceeds approximately 20, heat transport appears to become independent of system size. This observation implies that, for large system sizes, heat transport follows a ballistic pattern and deviates from Fourier's law. In contrast, for considerably smaller system sizes, the current exhibits a non-monotonic decline with $N$, attributed to finite size effects. Notably, this drop in heat current with $N$ is more pronounced in the topological phase compared to the trivial phase. { We suspect that the more pronounced drop of the heat current in the topological phase might be due to the presence of significant edge localisations, increasing the contribution from the edges. Separating the contribution from edge and bulk explicitly is not possible. So to gain a qualitative} understanding of this phenomenon, we employ two models to fit the behavior of heat current with system size. Both models are selected to ensure that heat current saturates in the thermodynamic limit, maintaining a ballistic transport.{ The bulk and edge current are defined such that the bulk current indicates the saturation current for large `$N$' and the edge current measures the decline in current with system size for smaller systems.}
	The first model, $F_1$, involves dividing the total current into ballistic and diffusive components \cite{Review_Modern_Physics}
	\begin{align}
		F_1:=		I_L(N)= J_B +J_E/N,
	\end{align}	
	where $J_B$ is the bulk current and $J_E$ is the edge current.
	The second model $F_2$ we use is given by
	\begin{align}
		F_2:=I_L(N)= J_B +J_E e^{-N}.
	\end{align}	
	{ Both these models ensure that for very small system sizes, the contribution from the edges to the total heat current remains  noticeable and as the system size increases the edge contribution becomes insignificant in comparison to the bulk.}
	Examining Fig. \ref{Edge_and_bulk_fig}(c), we observe that the fit is significantly better for the exponentially decaying model. However, in qualitative terms, both models reveal the same underlying property: as we enter the topological phases, the contribution from the edges increases rapidly, reaches a maximum, and then gradually declines.

	\par A practical approach to distinguish between edge and bulk current can be to isolate the current contributions originating from the two near-zero energy modes and compare them to the currents associated with energy modes at energies $\geq 2 \delta$. By varying the current contribution from energy modes below $2\delta$ with $N$ and defining the maximum contribution as $J_{Ep}$, we can draw insights from Fig. \ref{Edge_and_bulk_fig}(d), where we observe that this contribution qualitatively aligns with the fitting analysis we have presented.
	{ Based on our analysis of the total heat current, we suspect that the second derivative of the edge and the bulk currents might be an even better indicator of the phase transition. So we plot the curvature $I^{''}(\delta)$ as a function of $\delta$ in Fig. \ref{Edge_and_bulk_fig} (e) .  We see that the curvature has an extremum near the phase transition point ($\delta\sim 0$).  The phase transition point is the point of local maxima for the curvature of the edge current. Similarly it is a point of local minima for the curvature of the bulk current. From the graphs it appears that the curvature is a much better indicator for phase change in the edge and bulk current analysis, as it changes sign near the phase transition point $\delta =0$ .}
	

	\section{Effective temperature and steady state density matrix}\label{Sec:Effective_temp}
	The master equation \eqref{liouville} can be solved by defining an effective temperature \cite{nature_main_ref,EFF_temp_RevModPhys.93.041001,Eff_temp_PhysRevB.104.144301}
	\begin{align}	\label{effective_temperature}
		T^j_{eff}=\frac{|\omega_j|}{\log\Big{(}\frac{\kappa^L_je^{\frac{|\omega_j|}{T_L}}\big{(}e^{\frac{|\omega_j|}{T_R}}+1\big{)}+\kappa^R_je^{\frac{|\omega_j|}{T_R}}\big{(}e^{\frac{|\omega_j|}{T_L}}+1\big{)}}{\kappa^L_j\big{(}e^{\frac{|\omega_j|}{T_R}}+1\big{)}+\kappa^R_j\big{(}e^{\frac{|\omega_j|}{T_L}}+1\big{)}}\Big{)}}.
	\end{align}
	By defining this effective temperature, Eq. \eqref{liouville} reduces to,
	{
		\begin{align}
			\frac{d \hat{\rho}}{dt}&=-i[\hat{H}_S,\hat{\rho}]+\sum_{j} (\kappa^L_j +\kappa^R_j)  G(\omega_j,T^j_{eff})\mathcal{D}(\hat{a}_j)[\hat{\rho}]+(\kappa^L_j +\kappa^R_j) G(-\omega_j,T^j_{eff})\mathcal{D}(\hat{a}_j^{\dagger})[\hat{\rho}].
		\end{align}
	}%
	We can see that if $T_L=T_R$, $T_{eff}^j=T_L (T_R)$. Given that the global master equation ensures the thermalization of the density matrix at equilibrium, our system's steady state is consequently established as
	\begin{align}
		\hat{\rho}_{SS}=\frac{1}{Z}e^{-\sum_j {\omega_j}/{T_{eff}^j} \hat{a}_j^\dagger\hat{a}_j},
	\end{align}
	where the partition function is given by
	\begin{align}
		Z=Tr[e^{-\sum_j {\omega_j}/{T_{eff}^j} \hat{a}_j^\dagger\hat{a}_j}].
	\end{align}
	This indicates that we can conceptualize our system as an assembly of $N$ independent two-state systems, each in equilibrium with a bath at temperature $T_{eff}^j$. The energy gap between the states in each of these systems is determined by $\omega_j$.
	\begin{figure}[t]
		\centering 
		\subfigure []
		{\includegraphics[width=0.4\linewidth,height=0.3\linewidth]{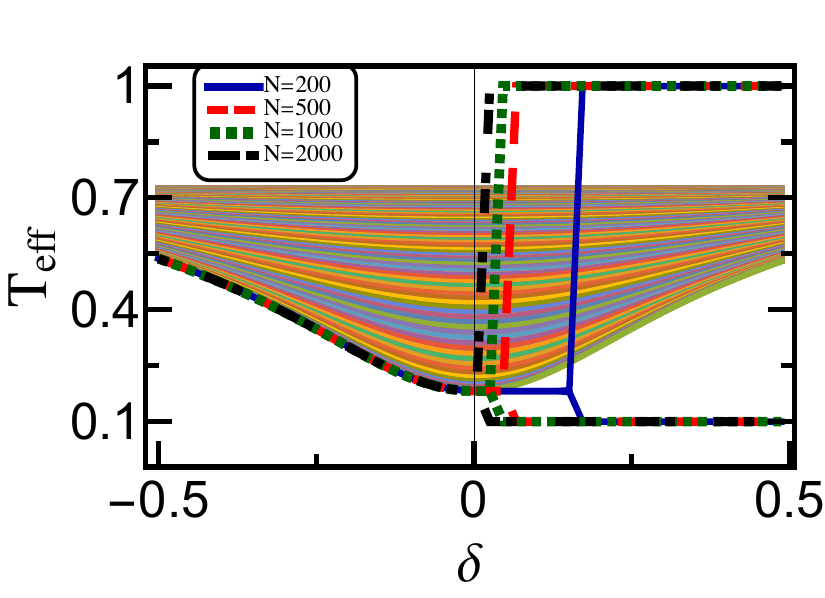}}
		\subfigure []
		{\includegraphics[width=0.4\linewidth,height=0.3\linewidth]{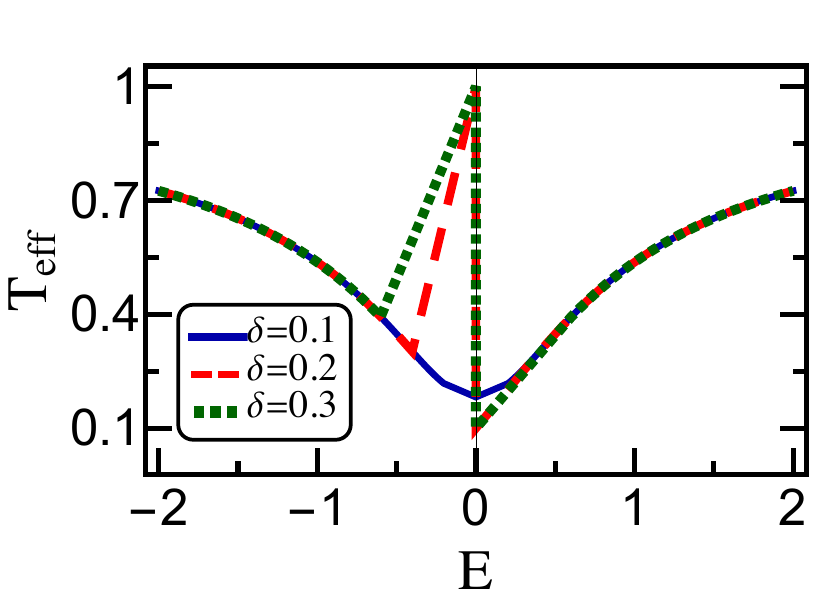}}
		\subfigure []
		{\includegraphics[width=0.4\linewidth,height=0.3\linewidth]{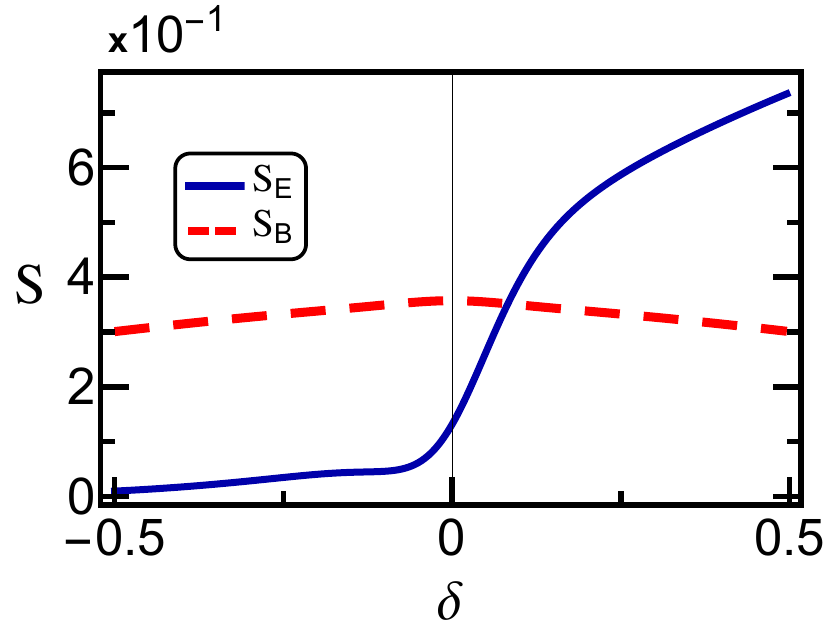}}
		\subfigure []
		{\includegraphics[width=0.4\linewidth,height=0.3\linewidth]{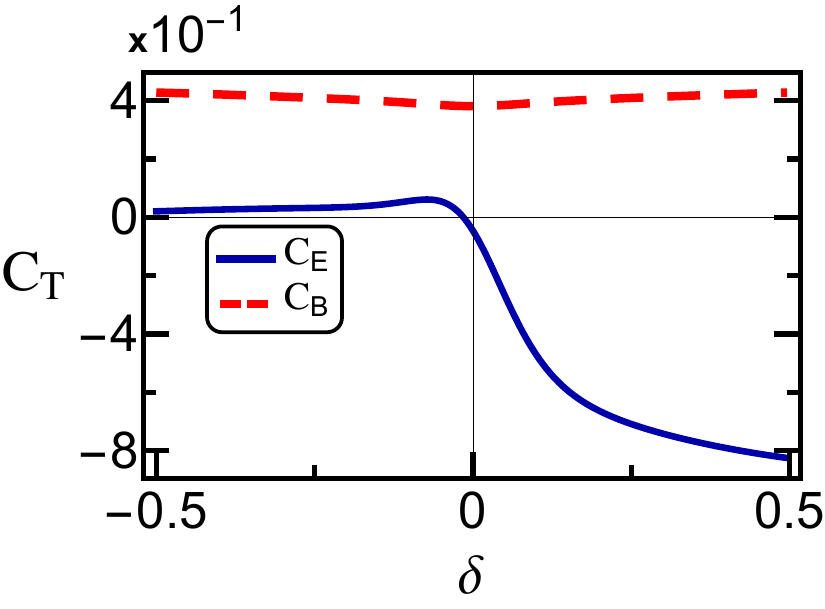}}
		\caption{{\textbf{(a)} Variation of the effective temperature spectrum with $\delta$. The near-zero energy mode effective temperature is also shown  for various system sizes $N$. }
			\textbf{(b)} Variation of effective temperature with energy $E$  for the Topological Phase. Variation of edge and bulk contribution with $\delta$ for \textbf{(c)} entropy $S$ and \textbf{(d)} specific heat $C_T$.}
		\label{effective_temp}
	\end{figure}
	\\\\
	Examining the behavior of $\kappa^L_j$ and $\kappa^R_j$ for the two near-zero energy modes, it becomes apparent that, owing to the localization of these modes at opposite ends, one tends toward zero while the other converges to one,
	\begin{align}
		\lim\limits_{\kappa^L_j (\kappa^R_j)\to1,\kappa^R_j (\kappa^L_j)\to0}T_{eff}^j=T_L (T_R).
	\end{align}
	When we apply this limit to the effective temperature expression in Eq. \eqref{effective_temperature}, we observe that one of the near-zero energy modes attains an effective temperature $T_L$, while the other reaches an effective temperature $T_R$. Additionally, given that $\kappa^L_j=\kappa^R_j$ for bulk energy modes, the effective temperature $T_{eff}^j$ for bulk modes remains unaffected by the wave function and relies solely on the energy of the modes as well as the temperatures of the left and right baths.
	
	{ Looking at the variation of the effective temperature spectrum with $\delta$ in Fig. \ref{effective_temp}(a), we see that while the temperature of bulk modes is same in both the phases, the effective temperatures of the two near zero energy modes separate out from the bulk spectrum in the topological phase. These two energy modes take the temperatures of the left or the right bath depending on where the wavefunction corresponding to them is localised. We also see that the separation of the two near-zero energy modes from the bulk occurs earlier in the topological phase for larger systems, as clearly seen in the figure.  In Fig. \ref{effective_temp}(b), we plot the effective temperature with energy of the modes in the topological phase. We see that the effective temperature of the non zero energy modes lie between the temperatures of the left and the right baths. Thus, we see a similar behavior when the $T_{eff}$ is plotted as a function of the energy of the modes in the topological phase. }
	
	\subsection{Thermodynamics of bulk and edge contributions}

	The effective partition function for our system can be written as
	\begin{align}
		Z=\Pi_{j=1}^{j=N} (1+e^{-\beta_j\omega_j}),
	\end{align} 
	where, $\beta_j=\frac{1}{T_{eff}^j}$. With this partition function, we can define the system's effective free energy
	\begin{align}\label{eq:free} 
		\mathcal{F}=-\sum_j\frac{1}{\beta_j}\log (1+e^{-\beta_j\omega_j}).=-\sum_j\mathcal{F}_j
	\end{align}
	Similarly, we can define the entropy of our system as
	\begin{align}
		S=-\sum_j\frac{\partial \mathcal{F}_j}{\partial T_{eff}^j}=\sum_j \log (1+e^{-\beta_j\omega_j})+\frac{\omega_j \beta_j e^{-{\beta_j \omega_j}}}{1+e^{-\beta_j\omega_j}}.
	\end{align}
	Utilizing the approach detailed in \cite{hill_kempkes_ref,Quelle_PhysRevB.94.075133}, we formulate the total entropy as a summation of edge entropy and bulk entropy
	\begin{align}
		S=S_E+S_B*N,
	\end{align}
	where $S_E$ and $S_B$ are edge and bulk entropies, respectively.
	Similarly, we can analyze the specific heat,
	\begin{align}
		C_T=\sum_j-T_{eff}^{j}\frac{\partial^2\mathcal{F}_j}{\partial T^{j2}_{eff}},
	\end{align}
	and using Eq. \eqref{eq:free}, the specific heat evaluates to
	\begin{align}
		C_T=\sum_j \frac{\beta_j^2 \omega_j^2e^{-\beta_j \omega_j} }{1+e^{-\beta_j\omega_j}}- \frac{\beta_j^2 \omega_j^2e^{-2\beta_j \omega_j} }{(1+e^{-\beta_j\omega_j})^2}.
	\end{align}
	Again distributing total specific heat into edge and bulk part using Hill's thermodynamics \cite{hill1994thermodynamics}, we have
	\begin{align}
		C_T=C_E+N*C_B
	\end{align}
	Upon performing data fitting analogously to \cite{hill_kempkes_ref}, we observe an excellent fit { because of the nano-thermodynamic support for the chosen functions}. As illustrated in Figs. \ref{effective_temp}(c) and \ref{effective_temp}(d), when transitioning from the trivial phase to the topological phase, there is a notable increase in the contribution from the edge to both entropy and specific heat. This observation underscores the heightened significance of the edge in the topological state, aligning with findings presented in \cite{hill_kempkes_ref} for an equilibrium system.

	
	\section{Heat diode with bosonic baths}\label{Sec:diode}
	It is reported in previous studies that in a spin-boson system, the asymmetry
	in the system bath coupling for a spin system sandwiched between two bosonic baths results in heat flow rectification \cite{dvira_segal_diode, PhysRevE.104.054137}. In our case, we observe from the heat current expression in Eq. \eqref{current_delta_figures} that, regardless of any type of asymmetry in system parameters, the heat current remains the same upon exchanging the bath temperatures.  { This is because just the structural asymmetry is not sufficient for inducing heat rectification in a system and some kind of non linearity or anharmonicity is required  \cite{dvira_segal_diode}. To overcome this, we opt for bosonic baths at the two ends instead of fermionic ones and reiterate our analysis. It has been shown in earlier studies \cite{Ozur_diff_stat_PhysRevE.106.054114, Diode:Type2_req_Dvira} that having a difference in particle statistics of the bath and system along with some structural asymmetry is sufficient for generating heat rectification. The system-bath interaction has the same form like the fermionic case and the expression for the current on doing this change is:}
	\begin{align}\label{boson_current}
		I_L=4\sum_{\omega_j>0}\omega_j  \frac{|\boldsymbol{\Phi}^j_{1 (N)}|^2\kappa_L \kappa_R\big{(}  N_L(\omega_j)- N_R(\omega_j)\big{)}}{\kappa_L(1+2 N_L(\omega_j))+\kappa_R(1+2 N_R(\omega_j))},
	\end{align}
	where,
	\begin{align}\label{Bose-Einstein}
		N_i(\omega)=\frac{1}{e^{\omega/T_i}-1}.
	\end{align}
	Once again, we note that the heat current is directly proportional to $|\boldsymbol{\Phi}^j_{1 (N)}|^2$. Consequently, the heat current exhibits a significant decrease in the topological phase.
	\subsection{Heat rectification}
	\begin{figure}[t]
		\centering 
		\subfigure []
		{\includegraphics[width=0.32\linewidth,height=0.25\linewidth]{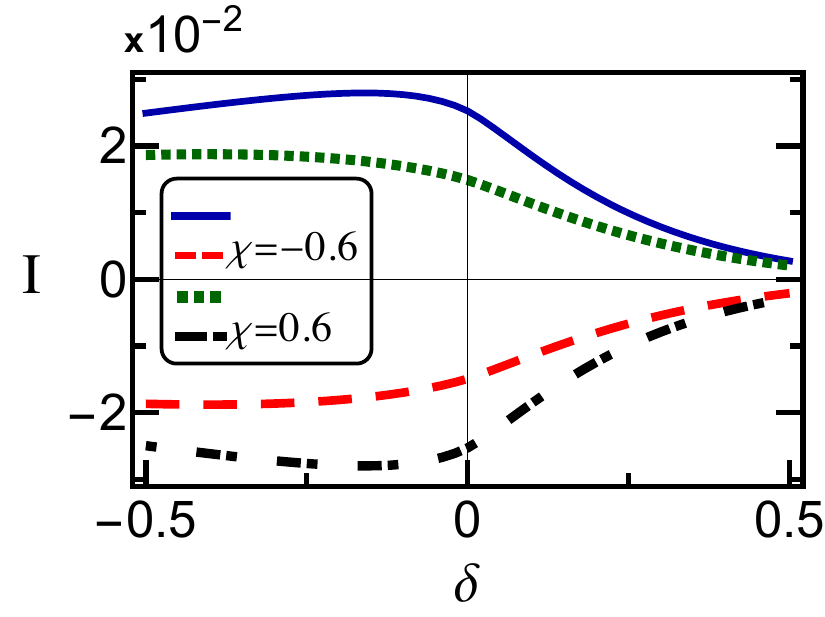}}
		\subfigure []
		{\includegraphics[width=0.32\linewidth,height=0.25\linewidth]{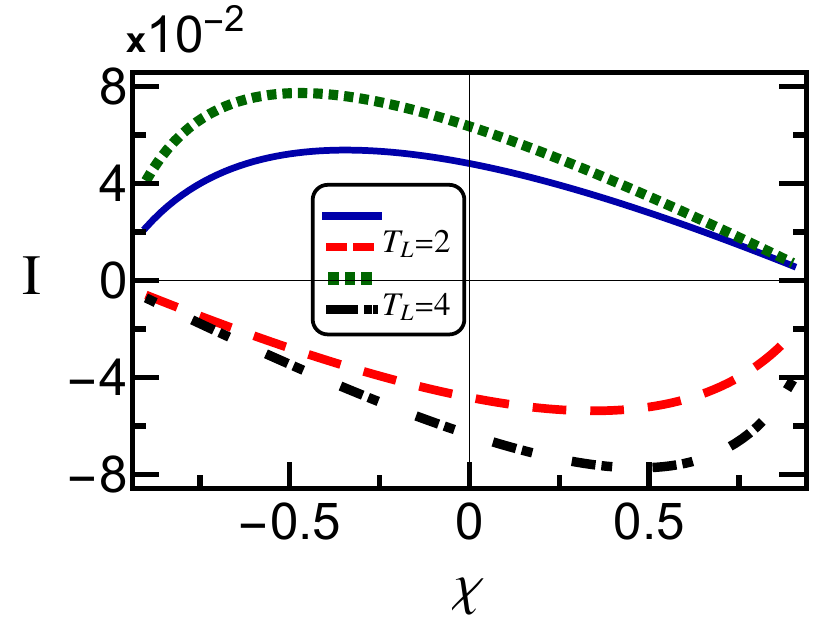}}
		\subfigure []
		{\includegraphics[width=0.32\linewidth,height=0.25\linewidth]{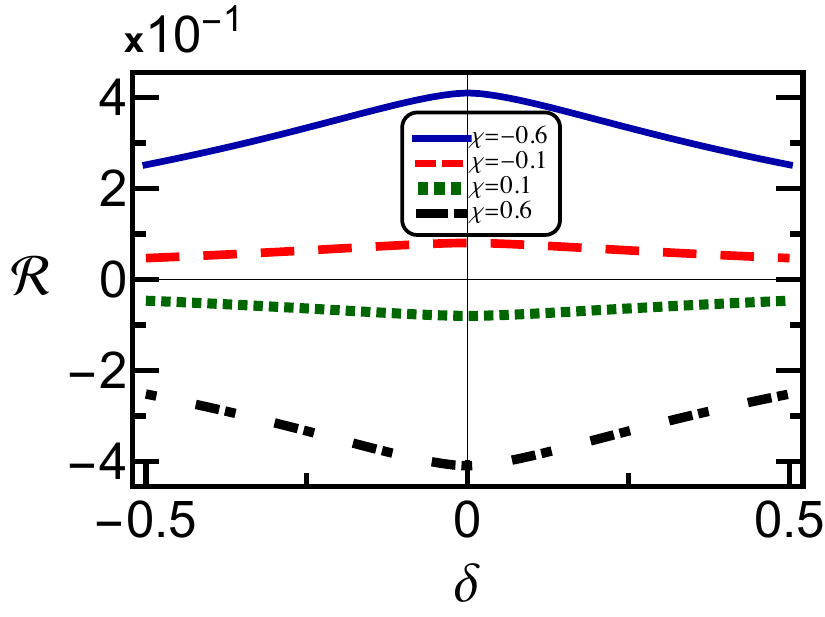}}
		\subfigure []
		{\includegraphics[width=0.32\linewidth,height=0.25\linewidth]{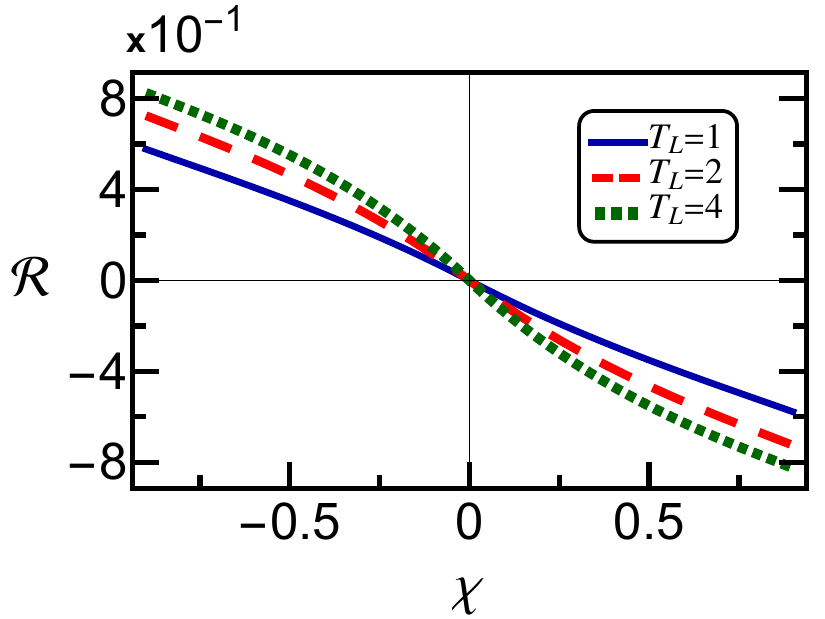}}
		\subfigure []
		{\includegraphics[width=0.32\linewidth,height=0.25\linewidth]{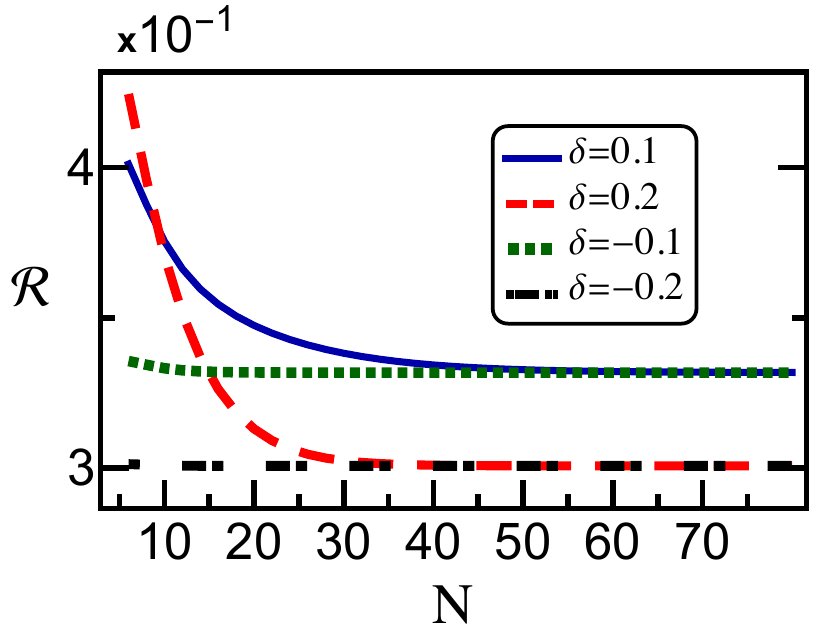}}
		\subfigure []
		{\includegraphics[width=0.32\linewidth,height=0.25\linewidth]{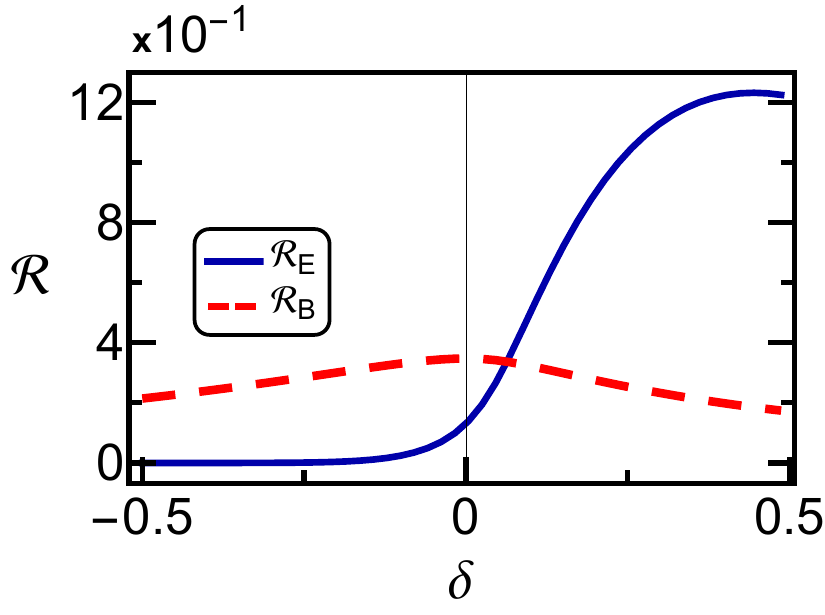}}
		\caption{Variation of left and right current with \textbf{(a)} $\delta$, \textbf{(b)} $\chi$, variation of rectification with \textbf{(c)} $\delta$, \textbf{(d)} $\chi$, \textbf{(e)}   system size $N$ and \textbf{(f)} Variation of edge and bulk contribution to rectification with $\delta$. If not specified otherwise $T_R=0.1, \chi=-0.5,\delta=0$. If the figures have empty legend titles, it signifies that we are showing both the left to right current and right to left current on exchanging temperatures for the same value of other parameters. The sign of current is positive for left to right current and vice versa. This means that (blue continuous line) and (green dotted line) represent the left to right currents and (red dashed line) and (black dot-dashed line) represent the right to left currents.}
		\label{rectification_fig}
	\end{figure}
	
	When we exchange the temperatures of the bath in equation Eq. \eqref{boson_current}, the denominator undergoes a change  
	\cite{dvira_segal_diode}, while the numerator remains constant. This implies the potential for heat rectification when $\kappa_L\ne\kappa_R$. Considering this, we assume 
	\begin{align}
		&\kappa_L=\kappa(1+\chi),&&\kappa_R=\kappa(1-\chi),&&&\kappa=0.1.
	\end{align}
	The difference in heat current on the exchange of temperature is given by \cite{dvira_segal_diode}
	\begin{align}
		\Delta I&=I_L(T_L,T_R)+I_L(T_R,T_L)\nonumber\\&=\sum_{\omega_j>0}  \frac{-4\omega_j|\boldsymbol{\Phi}^j_{1 (N)}|^2 \kappa\chi(1-\chi^2)(N_L(\omega_j)-N_R(\omega_j))^2}{(1+N_L(\omega_j)+N_R(\omega_j))^2-\chi^2(N_L(\omega_j)-N_R(\omega_j))^2}.
	\end{align}
	\textbf{Sign Convention:} We adopt a positive sign convention for the left-to-right bath current for $T_L>T_R$, and vice versa. This convention also dictates the sign of rectification, with ``positive" indicating that the current flows more from left to right, and vice versa. The rectification factor can be calculated by \cite{diode4, PhysRevE.104.054137}
	\begin{equation}\label{eq:rectification}
		\mathcal{R}=\frac{ I_L(T_L,T_R)+I_L(T_R,T_L)}{\text{Max}[| I_L(T_L,T_R)|,| I_L(T_R,T_L) |]}.
	\end{equation}
	We observe that the current is suppressed from left to right when the left bath is more strongly connected to the system, and vice versa.
	
	In Fig. \ref{rectification_fig} (a), we observe that the behavior of heat current with respect to the topological order parameter $\delta$ is qualitatively similar to the fermionic case. There is a sharp decline in the heat flow as we transition from the trivial phase into the topological phase. This decline is attributed to the heat flow dependency on the value of $|\boldsymbol{\Phi}^j_{1 (N)}|^2$. Analyzing Fig. \ref{rectification_fig}(b), we observe that the introduction of coupling strength asymmetry results in the suppression of current from the strongly coupled site, consistent with the findings of previous studies \cite{dvira_segal_diode}. Examining the rectification in relation to $\delta$ in Fig. \ref{rectification_fig}(c), we find that based on the asymmetry of heat flow it is not possible to differentiate between the trivial and topological phases. Lastly, Fig. \ref{rectification_fig}(d) reveals that reversing the sign of the asymmetry parameter $\chi$ results in a reversal of the current rectification direction.

	\begin{figure}[t]
		\centering 
		\subfigure []
		{\includegraphics[width=0.32\linewidth,height=0.25\linewidth]{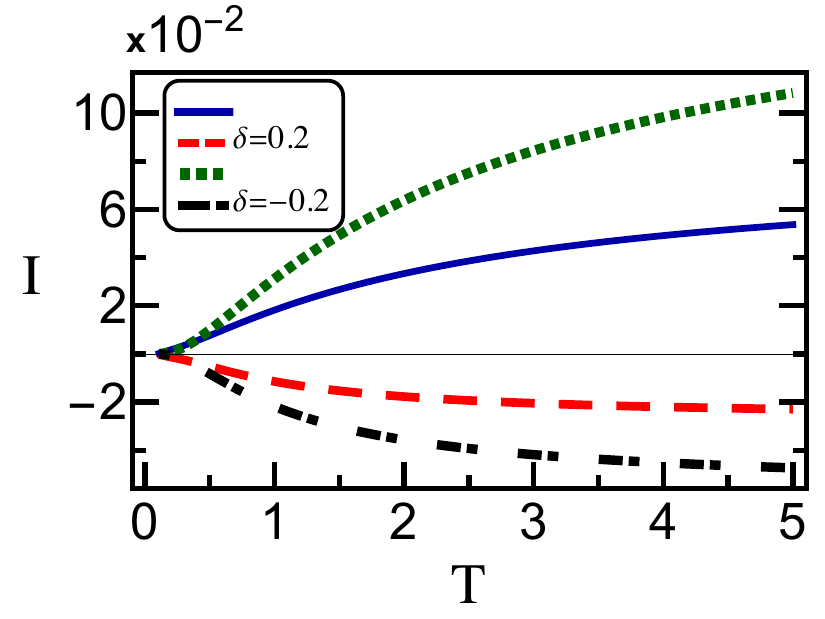}}
		\subfigure []
		{\includegraphics[width=0.32\linewidth,height=0.25\linewidth]{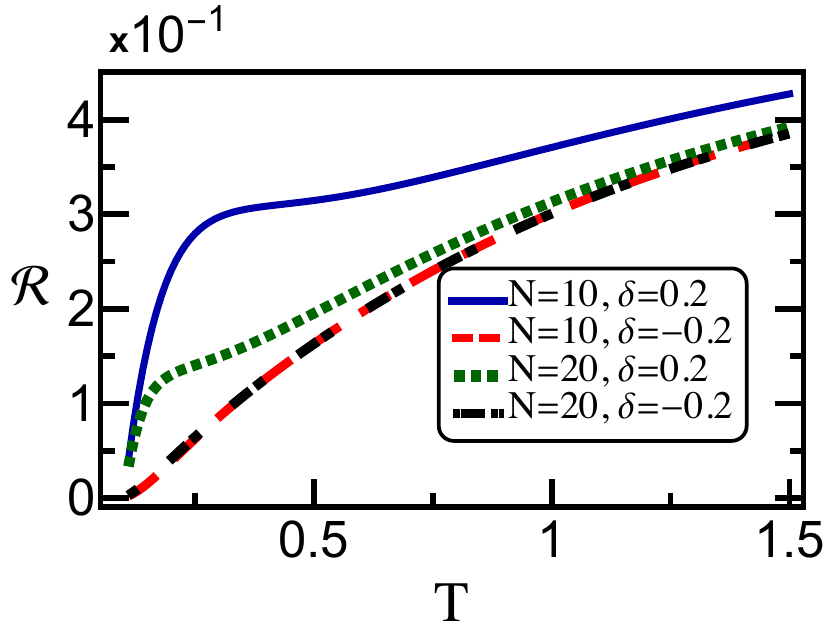}}
		\subfigure []{\includegraphics[width=0.32\linewidth,height=0.25\linewidth]{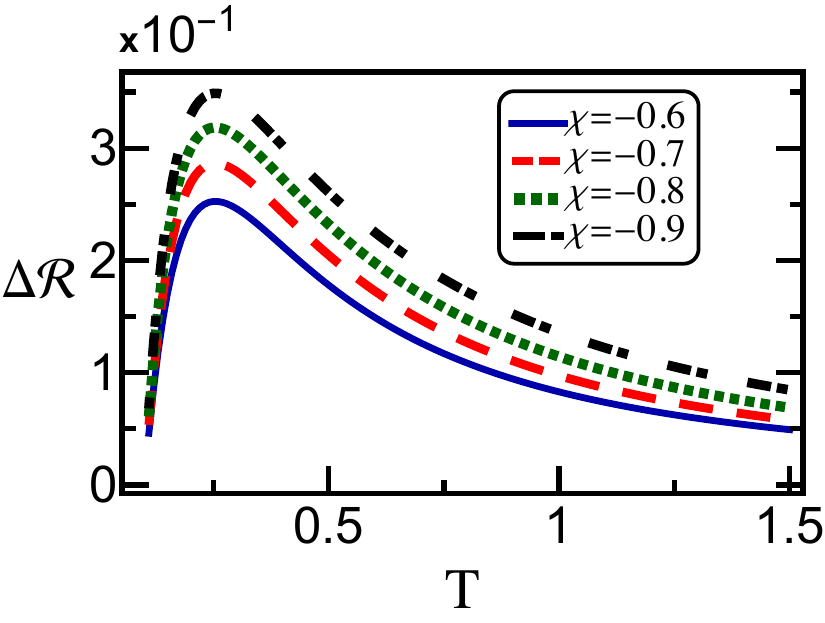}}
		\subfigure []{\includegraphics[width=0.32\linewidth,height=0.25\linewidth]{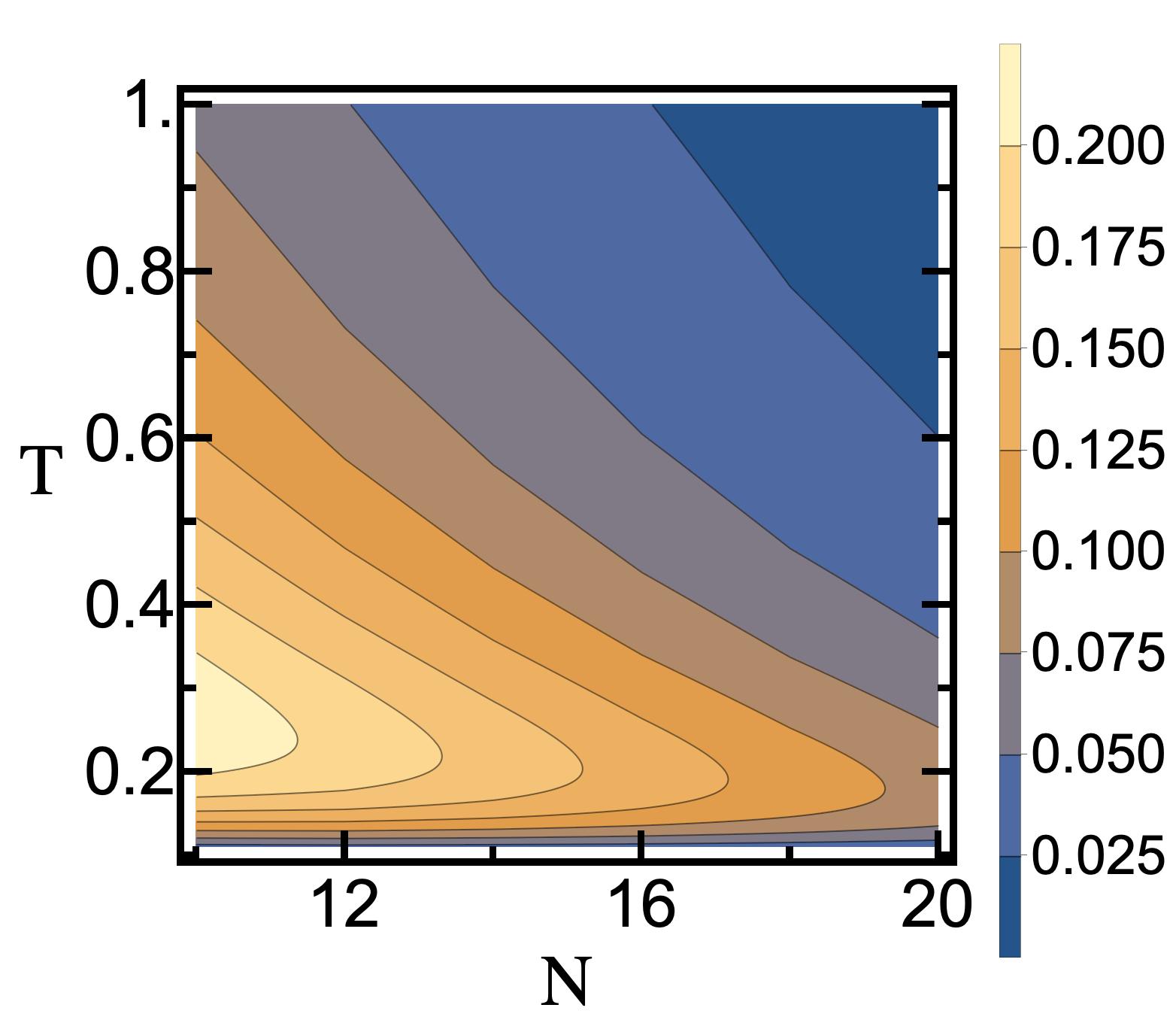}}
		\subfigure []{\includegraphics[width=0.32\linewidth,height=0.25\linewidth]{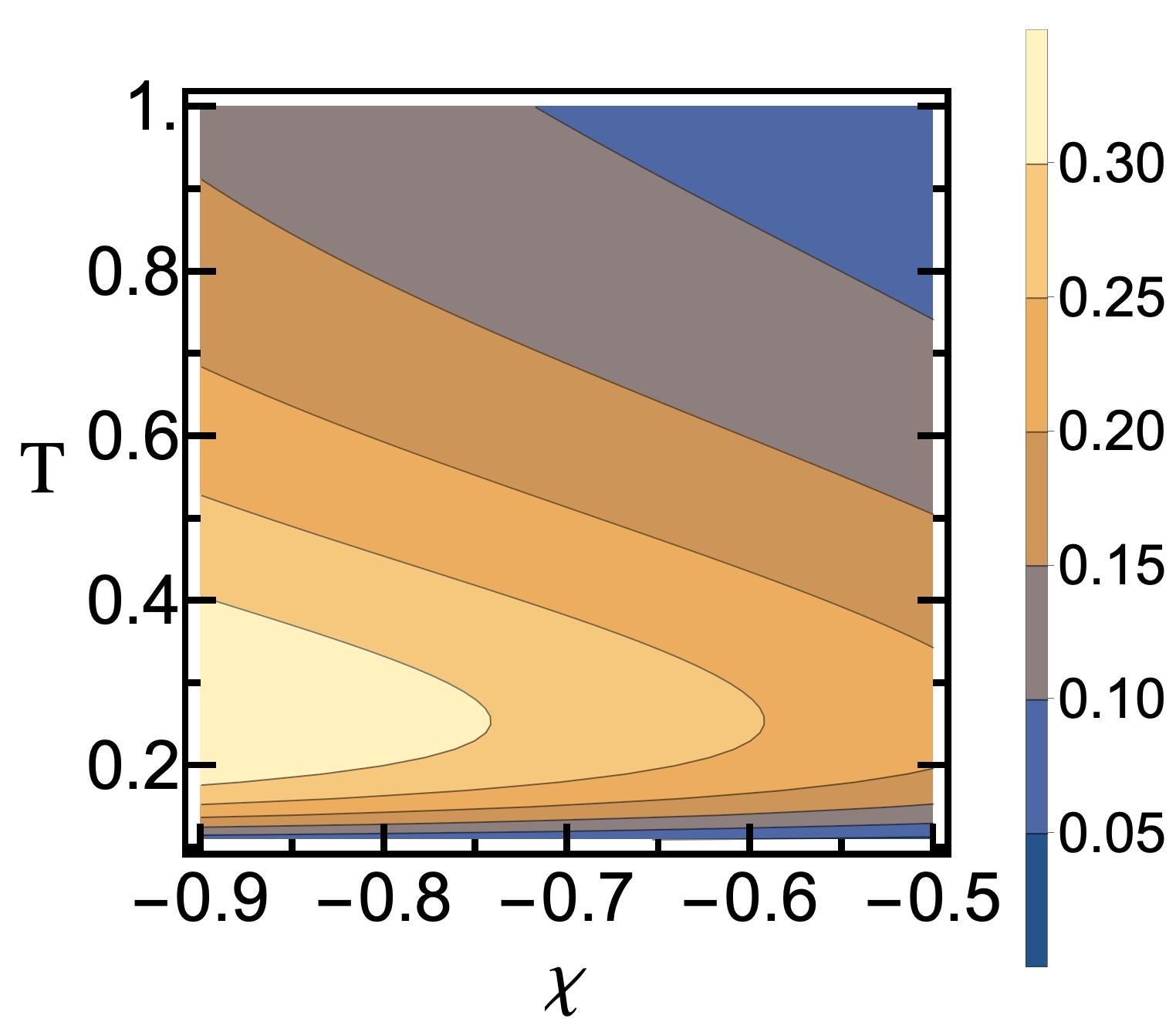}}
		\subfigure []{\includegraphics[width=0.32\linewidth,height=0.25\linewidth]{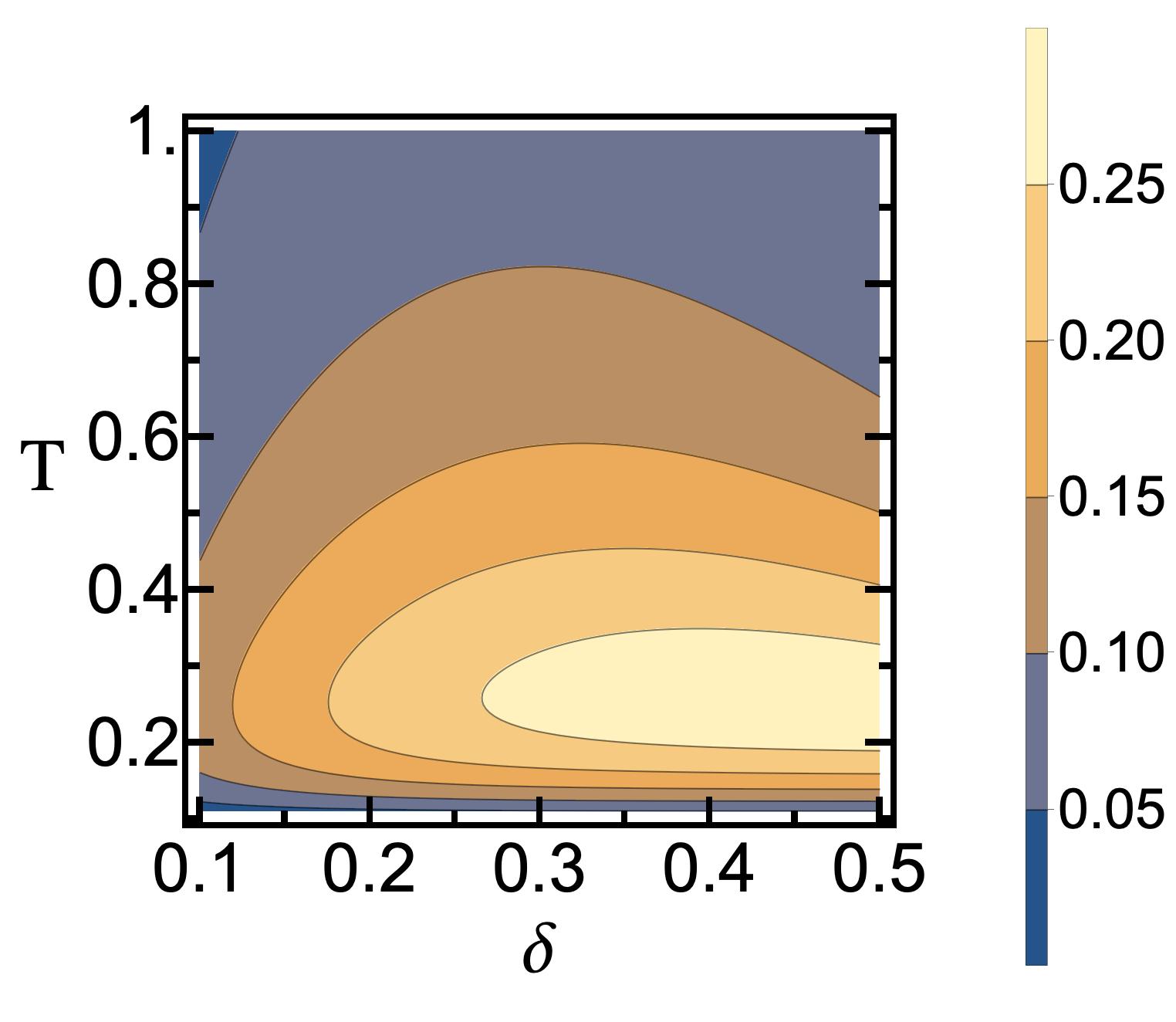}}
		\caption{\textbf{(a)} Variation of left and right current with Temperature $T$ for trivial and topological phase, \textbf{(b)} Variation of Rectification with Temperature $T$, \textbf{(c)} Variation of difference between rectification in topological phase and trivial phase ($\Delta \mathcal{R}$) with $T$. Contour plots showing variation of $\Delta \mathcal{R}$ with \textbf{(d)} $N, T$, \textbf{(e)} $\chi, T$ and \textbf{(f)} $\delta,T$. While doing the variations, temperature of one bath is fixed at $`T_L (T_R)=0.1'$. The parameters if not otherwise specified are `$\chi=-0.5, N=10$ and $\delta=0.2$'.}
		\label{fig_few_sites}
	\end{figure}
	
	\subsection{Diode with a few fermionic sites}
	The preceding analysis indicates that, for large systems, our diode's rectification ability is not a robust indicator for distinguishing between the topological and trivial phases. However, when we plot rectification against system size in Fig. \ref{rectification_fig}(e), a consistent pattern emerges for small $N$: rectification is consistently greater in the topological phase. {As we incrementally increase the system size, the rectification becomes equal in both phases. This is because for large systems sizes, the contribution of the edge current to the total current becomes insignificant.}
	Accordingly, for SSH model with few fermionic sites, our diode proves to be a reliable indicator of the system's topological phase. When examining the rectification curve with respect to system size $N$, we again make a fit using the following function
	\begin{align}
		\mathcal{R}=\mathcal{R}_B+\frac{\mathcal{R}_E}{N}.
	\end{align}
	We observe a sharp increase in the contribution from the edge as we enter the topological phase. Similar to the heat flow analysis concerning bulk and edge contributions (as shown in Fig. \ref{Edge_and_bulk_fig}), we differentiate between the heat flow linked to near-zero energy modes and that associated with modes at energies $>2 \delta$. We plot the contribution of edge and bulk heat flow asymmetry in the rectification factor $\mathcal{R}$ as a function of the order parameter $\delta$ in Fig. \ref{rectification_fig}(f). It becomes evident from the results that in the trivial phase, only bulk heat flow contributes to heat rectification. In contrast, in the topological phase, the contribution of edge heat flow asymmetry dominates over bulk heat flow. This observation could serve as an indicator for distinguishing between the trivial and topological phases in the SSH model. 
	{To better understand the ability of diode in distinguishing trivial and topological phase, we look at the plots in Fig. \ref{fig_few_sites}. In Fig. \ref{fig_few_sites}(a), we plot the variation of currents with temperature and observe that the heat current increases non-monotonically with temperature in both the trivial and topological phase before saturating for large temperature gradients. On plotting the corresponding rectification with temperature in Fig. \ref{fig_few_sites}(b), we observe that for small temperature gradient the difference in rectification between topological phase and trivial phase is large. Since for these cases the rectification in the trivial phase is almost zero, the phase differentiation ability of the diode is much higher for these temperature gradients values.
		Finally, we define $`\Delta \mathcal{R}=\mathcal{R}_{\delta}-\mathcal{R}_{-\delta}$, as the parameter which measures the capacity of diode to differentiate between topological and trivial phase. Plotting the contour plots of $\Delta \mathcal{R}$ with various system parameters in \ref{fig_few_sites}(d), \ref{fig_few_sites}(e) and \ref{fig_few_sites}(f), we observe that the optimal condition for the diode to be used as a topological phase identifier is when the system has few fermionic sites N, the temperature gradient between the baths is not very large, the system is away from the phase transition point $\delta=0$ and the bath coupling asymmetry parameter $\chi$ is large. Remarkably, all these conditions lead to low value for heat current between the baths. Considering all this, we can also conclude that the rectification ability is a good indicator of  phase for systems with few fermionic sites when the heat flow in the system is small.}

	\section{Conclusions}\label{sec:conclusion}
	In this work, we have explored the potential of thermodynamics as a tool for identifying topological phase transitions in the one-dimensional Su–Schrieffer–Heeger (SSH) model. Our investigation revealed intriguing thermodynamic signatures associated with the transition from the trivial to the topological phase, emphasizing the unique transport properties that emerge in the SSH model. The thermodynamically consistent global master equation served as a powerful framework for our analysis, allowing us to precisely quantify the suppression of heat flow as we enter the topological phase. This decline in heat flow was linked to the reduction in transmission coefficients of non-zero energy modes within the topological phase. We have demonstrated that this reduction can be a valuable indicator of a phase transition in the SSH model. Moreover, our investigation of heat flow asymmetry provided additional insights. We observed that when employing fermionic baths, no asymmetry was detected. However, upon substituting fermionic baths with bosonic ones, a non-zero heat flow asymmetry emerged. Remarkably for systems with few fermionic sites, this asymmetry was more pronounced in the topological phase compared to the trivial phase. Consequently, for such systems the behavior of the heat diode serves as an additional means of distinguishing between these two phases. While such asymmetry has been used for rectification in earlier systems, its role as an indicator of phase transition in SSH chains has not been thoroughly studied before. Our findings contribute to the broader understanding of the intriguing behavior exhibited by the SSH model, providing insights that may have implications for various fields, including condensed matter physics and quantum information science.

	\section{Acknowledgements}
	RM gratefully acknowledges financial support from Science and Engineering Research Board (SERB), India, under
	the Core Research Grant (Project No. CRG/2020/000620).
	\"O.~E.~M.~gratefully acknowledges Faculty of Engineering and Natural Sciences of Sabanc\i ~University
	for their hospitality during his sabbatical.
	\appendix
	\section{Master Equation}\label{section:Appendix A}
	The master equation derivation is similar to the one given in \cite{nature_main_ref}.
	We begin by writing the total Hamiltonian,
	\begin{align}\label{system_Hamiltonian}
		\hat{H}_T=\hat{H}_S+\hat{H}_B+\hat{H}_{SB}
	\end{align}
	where the system Hamiltonian is,
	\begin{align}
		\hat{H}_S=\hat{V}+\hat{H}_0=\langle C^{\dagger}| M | C\rangle
	\end{align}
	where $M$ is a matrix and
	$|C \rangle=(|\hat{c}_1,\hat{c}_2,\hat{c}_3,\hat{c}_4\ldots \hat{c}_N\rangle)^T$.
	We can diagonalise the system Hamiltonian using the unitary transformation:
	\begin{align}
		M_D&=U^{\dagger}MU, &&
		M=U M_D U^{\dagger},
		&&M_{i,j}=\sum_{k,l}U_{i,l} (M_D)_{l,k}U^{\dagger}_{k,j}=\sum_{k}U_{i,k} (M_D)_{k,k}U^{\dagger}_{k,j}
	\end{align}
	So,
	\begin{align}
		H_S&=\langle C^{\dagger}| M | C\rangle=\sum_{i,j} \hat{c}^{\dagger}_i M_{i,j} \hat{c}_j =\sum_{i,j,k} \hat{c}^{\dagger}_iU_{i,k} (M_D)_{k,k}U^{\dagger}_{k,j}\hat{c}_j
	\end{align}
	
	If,
	\begin{align}\label{operator_relation}
		\hat{a}_k&=\sum_j U^\dagger_{k,j} \hat{c}_j ,&&
		\hat{a}^{\dagger}_k=\sum_j U_{j,k} \hat{c}^{\dagger}_j,  &&&|A\rangle =U^{\dagger} |C \rangle, 
		&&&&|C\rangle =U|A \rangle
	\end{align}
	Studying the anti-commutation relation between $\hat{a}'s$. Since,
	\begin{align}
		&	\{\hat{c}_i,\hat{c}_j\}=0,&&	\{\hat{c}^{\dagger}_i,\hat{c}^{\dagger}_j\}=0,&&	\{\hat{c}_i,\hat{c}^{\dagger}_j\}=\delta_{i,j}.
	\end{align}
	We can show,
	\begin{align}
		&	\{\hat{a}_i,\hat{a}_j\}=0,&&	\{\hat{a}^{\dagger}_i,\hat{a}^{\dagger}_j\}=0,&&	\{\hat{a}_i,\hat{a}^{\dagger}_j\}=\delta_{i,j}.
	\end{align}
	So, we can write the system Hamiltonian in the energy basis as:
	\begin{align}
		H_S=\sum_{k} (M_D)_{k,k} \hat{a}^{\dagger}_k \hat{a}_k=\sum_k \omega_k\hat{a}^{\dagger}_k \hat{a}_k
	\end{align}
	where, $\omega_k$ is the energy of the $k^{th}$ energy mode. Writing the system bath interaction Hamiltonian in the diagonal basis of the system Hamiltonian. Since, $ \hat{c}_1=\sum_{k}U_{1,k}\hat{a}_k,\hat{c}_1^{\dagger}=\sum_{k}U^{\dagger}_{k,1}\hat{a}^{\dagger}_k$
	So,
	\begin{align}
		H^L_{SB}=(\hat{c}_1^{\dagger}+\hat{c}_1)\otimes \sum_k (g^L_k \hat{b}_k+g^L_k\hat{b}_k^{\dagger})
		=\sum_j(U_{1,j}\hat{a}_j+U^{\dagger}_{j,1}\hat{a}^{\dagger}_j)\otimes \sum_k (g^L_k \hat{b}^{\dagger L}_k+ g^{*L}_k \hat{b}^L_k )=\sum_j\hat{X}^L_j\otimes\hat{Y}^L_j.
	\end{align}
	where,
	\begin{align}
		&\hat{X}^L_j=U_{1,j}\hat{a}_j+U^{\dagger}_{j,1}\hat{a}^{\dagger}_j,&&\hat{Y}^L_j=\hat{Y}_L= \sum_{k}g^L_k \hat{b}^{\dagger L}_k+ g^{*L}_k \hat{b}^L_k 
	\end{align}
	
	The Redfield master equation in the interaction picture is given as \cite{breuer2002},
	
	\begin{equation}\label{BornMarkov}
		\frac{d \hat{\rho}^I(t)}{dt}=-\int_{0}^{\infty}ds\text{   } Tr_B\{[\hat{H}_{SB}^{I}(t),[\hat{H}_{SB}^{I}(t-s),\hat{\rho}_S^I(t)\otimes\hat{\rho}_B^I(0)]]\}
	\end{equation}
	
	Simplifying,
	\begin{align}
		\frac{d \hat{\rho}^I(t)}{dt}=\mathcal{D}_L[\hat{\rho}]+\mathcal{D}_R[\hat{\rho}]
		=\sum_{k,l}\int_{0}^{\infty} d\tau \Gamma_{l,k}(\tau,T) [\hat{X}_k(t-\tau)\hat{\rho},\hat{X}_l(t)]+H.C.
	\end{align}
	and,
	\begin{align}
		\Gamma_{k,l}(\tau,T)=Tr_{B}(\hat{\rho}_B \hat{Y}_k(\tau) \hat{Y}_l)
	\end{align}

	{Finding the left dissipator terms,	since, $\hat{Y}_j=\hat{Y}_L =\sum_{k}g^L_k \hat{b}^{\dagger L}_k+ g^{*L}_k \hat{b}^L_k $,
		\begin{align}
			\Gamma_{k,l}(\tau,T)=	\Gamma(\tau,T)=Tr_{B}(\hat{\rho}_B \hat{Y}_L(\tau) \hat{Y}_L)\nonumber &=\sum_k|g^L_k|^2[e^{i \omega^L_k} f_L({\omega^L_k})+e^{-i \omega^L_k} (1-f_L(\omega^L_k))]
			\nonumber\\&=\int_{0}^{\infty}d\omega \textnormal{J}^L(\omega)[e^{i \omega} f_L({\omega})+e^{-i \omega} (1-f_L(\omega))]		
		\end{align}
		where the bath spectrum function is defined as,
		\begin{align}\label{J_definition}
			\textnormal{J}^L(\omega)=\sum_k |g^L_k|^2 \delta(\omega-\omega^L_k)=\kappa_L J^L(\omega)
		\end{align}
		In this study, we will assume that both the baths have same functional part, so $J^L(\omega)=J^R(\omega)=J(\omega)$ though the coupling strength $\kappa_{L(R)}$ may be different.}
	So,
	\begin{align}
		\mathcal{D}_L[\hat{\rho}]&=\sum_k \int_{0}^{\infty} d\tau \Gamma(\tau,T)  [\hat{X}^L_k(t-\tau)\hat{\rho},\hat{X}^L_l(t)]+H.C. \nonumber\\
		&=\sum_k |U_{1,k}|^2\int_{0}^{\infty} d\tau \Gamma(\tau,T) e^{i \tau \omega_k}[\hat{a}_k\hat{\rho},\hat{a}^\dagger_k]+H.C. +\sum_k|U_{1,k}|^2 \int_{0}^{\infty} d\tau \Gamma(\tau,T)e^{-i \tau \omega_k}  [\hat{a}^\dagger_k\hat{\rho},\hat{a}_k]+H.C. 
		\	\end{align}
	Putting in the value of $\Gamma(\tau,T)$ and using the secular approximation,
	we get the master equation given in Eq. \eqref{main_eqs} in the Schr\"{o}dinger picture \cite{breuer2002}.
	
	\begin{align}\label{main_eqs}
		\frac{d \hat{\rho}}{dt}&=-i[\hat{H}_S,\hat{\rho}]+\sum_{j} \Big{(}\kappa^L_j G(\omega_j,T_L)+\kappa^R_j G(\omega_j,T_R)\Big{)} \mathcal{D}(\hat{a}_j)[\hat{\rho}]+\Big{(}\kappa^L_j G(-\omega_j,T_L)+\kappa^R_j G(-\omega_j,T_R)\Big{)}\mathcal{D}(\hat{a}_j^{\dagger})[\hat{\rho}]
	\end{align}

	\section{Heat Current Expression} \label{Appendix:B}
	
	The heat current,
	\begin{align} \label{Appendix_heat}
		I_L&=Tr(\mathcal{L}_L H_S)= 2\sum_{j}-\omega_j \kappa^L_j G(\omega_j,T_L) \langle\hat{a}^\dagger_j \hat{a}_j \rangle +\omega_j\kappa^L_j G(-\omega_j,T_L) (1-\langle\hat{a}^\dagger_j \hat{a}_j \rangle) 
		=2\sum_{j}\frac{|\omega_j| (k^L_jk^R_j)}{k^L_j+k^R_j}{J(|\omega_j|)}(F_L(|\omega_j|)-F_R(|\omega_j|))	 
	\end{align}
	The value of $\langle \hat{a}_k^{\dagger}\hat{a}_k \rangle$ in the above equation has been substituted  from the following master equation  \eqref{main_eqs}, we get,
	\begin{align}
		\frac{d \langle \hat{a}^{\dagger}_k\hat{a}_k\hat{\rho}\rangle}{dt}&=-i \langle \hat{a}^{\dagger}_k\hat{a}_k[\hat{H}_S,\hat{\rho}]\rangle\nonumber+\Big{\langle}\sum_{j}\Big{(}\kappa^L_j G(\omega_j,T_L)+\kappa^R_j G(\omega_j,T_R)\Big{)}\hat{a}^{\dagger}_k\hat{a}_k\mathcal{D}(\hat{a}_j)[\hat{\rho}]\nonumber\\&+\Big{(}\kappa^L_j G(-\omega_j,T_L)+\kappa^R_jG(-\omega_j,T_R)\Big{)}\hat{a}^{\dagger}_k\hat{a}_k\mathcal{D}(\hat{a}_j^{\dagger})[\hat{\rho}]\Big{\rangle}
	\end{align}
	
	Solving for the steady state,
	\begin{align}
		&\langle \hat{a}^{\dagger}_k\hat{a}_k \rangle={\frac{\Big{(}\kappa^L_j G(-\omega_j,T_L)+\kappa^R_j G(-\omega_j,T_R)\Big{)}}{\Big{(}\kappa^L_j G(\omega_j,T_L)+\kappa^R_j G(\omega_j,T_R)\Big{)}+\Big{(}\kappa^L_j G(-\omega_j,T_L)+\kappa^R_jG(-\omega_j,T_R)\Big{)}}} =\frac{\Big{(}\kappa^L_j G(-\omega_j,T_L)+\kappa^R_j G(-\omega_j,T_R)\Big{)}}{\kappa^L_j +\kappa^R_j}
	\end{align}
	\section{Heat Current for different bath spectrum functions} \label{Appendix:C}
	
	\renewcommand\thefigure{\thesection.\arabic{figure}}   
	\setcounter{figure}{0}   
	\begin{figure}[!htbp]
		\centering 
		\subfigure []
		{\includegraphics[width=0.32\linewidth,height=0.25\linewidth]{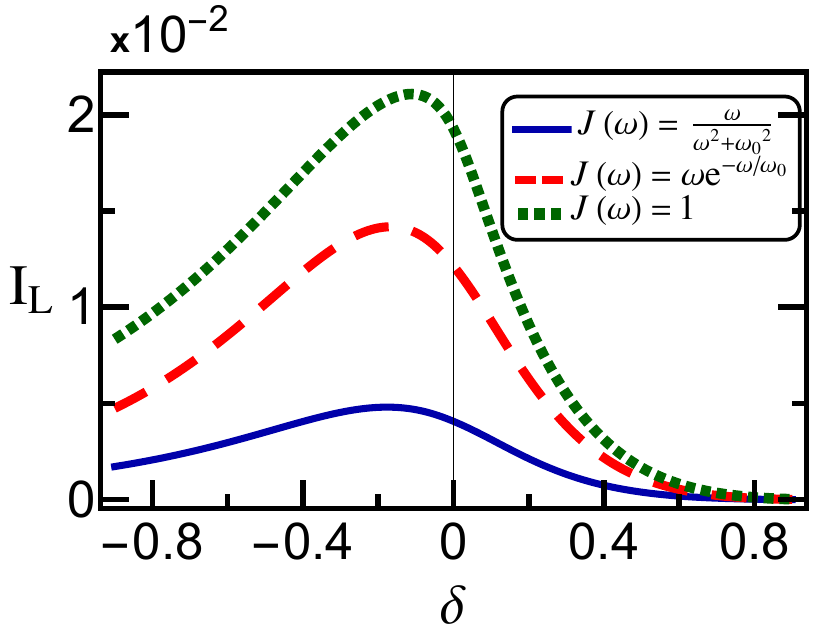}}
		\subfigure[]
		{\includegraphics[width=0.32\linewidth,height=0.25\linewidth]{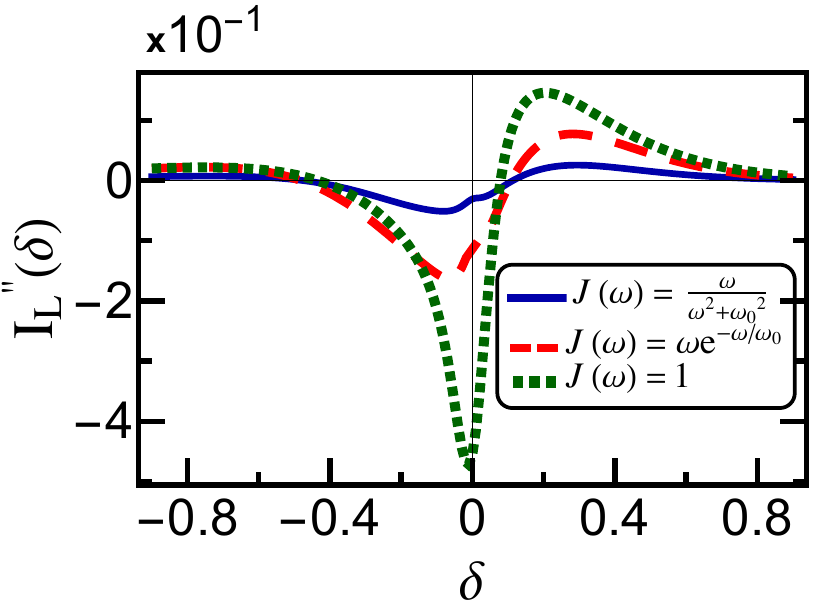}}
		\caption{{\textbf{(a)} Variation of heat current with $\delta$ for various bath spectrum functions and  and \textbf{(b)} Variation of the curvature of the heat current with $\delta$. For all the case, if not specified otherwise, the values of the parameters are $N=200, \omega_0=2, t=2, T_L=1, T_R=0.1, \delta=0.2, \kappa_L=0.1, \kappa_R=0.1$.}}
		\label{Appendix_current_delta_figures}
	\end{figure}
	{
		Though we work with the flat spectrum in the main text, the choice of the bath spectrum does not alter the qualitative behavior of the heat current. This is because the bath spectrum function changes just a multiplication factor of the heat current in Eq. \eqref{Appendix_heat}. To show this, we plot the heat current with the Rubin's bath spectrum $J(\omega)=\frac{\omega}{\omega+\omega_0^2}$ and a bath of the form $J(\omega)=\omega e^{-\omega/\omega_0}$ along with the flat spectrum. Here $\omega_0$ indicates the cutoff frequency of the bath and any contribution from frequencies greater than $\omega_0$ is suppressed.  We see from Fig. \ref{Appendix_current_delta_figures} (a) and (b) that the qualitative behavior of the heat current with different bath spectrum functions remains the same. The heat current suppresses in the topological phase and the curvature of the heat current indicates the phase transition point for any choice of bath spectrum.}
	
	
	
	\bibliography{SSH}

\providecommand{\noopsort}[1]{}\providecommand{\singleletter}[1]{#1}%
\begin{thebibliography}{51}%
\makeatletter
\providecommand \@ifxundefined [1]{%
 \@ifx{#1\undefined}
}%
\providecommand \@ifnum [1]{%
 \ifnum #1\expandafter \@firstoftwo
 \else \expandafter \@secondoftwo
 \fi
}%
\providecommand \@ifx [1]{%
 \ifx #1\expandafter \@firstoftwo
 \else \expandafter \@secondoftwo
 \fi
}%
\providecommand \natexlab [1]{#1}%
\providecommand \enquote  [1]{``#1''}%
\providecommand \bibnamefont  [1]{#1}%
\providecommand \bibfnamefont [1]{#1}%
\providecommand \citenamefont [1]{#1}%
\providecommand \href@noop [0]{\@secondoftwo}%
\providecommand \href [0]{\begingroup \@sanitize@url \@href}%
\providecommand \@href[1]{\@@startlink{#1}\@@href}%
\providecommand \@@href[1]{\endgroup#1\@@endlink}%
\providecommand \@sanitize@url [0]{\catcode `\\12\catcode `\$12\catcode
  `\&12\catcode `\#12\catcode `\^12\catcode `\_12\catcode `\%12\relax}%
\providecommand \@@startlink[1]{}%
\providecommand \@@endlink[0]{}%
\providecommand \url  [0]{\begingroup\@sanitize@url \@url }%
\providecommand \@url [1]{\endgroup\@href {#1}{\urlprefix }}%
\providecommand \urlprefix  [0]{URL }%
\providecommand \Eprint [0]{\href }%
\providecommand \doibase [0]{http://dx.doi.org/}%
\providecommand \selectlanguage [0]{\@gobble}%
\providecommand \bibinfo  [0]{\@secondoftwo}%
\providecommand \bibfield  [0]{\@secondoftwo}%
\providecommand \translation [1]{[#1]}%
\providecommand \BibitemOpen [0]{}%
\providecommand \bibitemStop [0]{}%
\providecommand \bibitemNoStop [0]{.\EOS\space}%
\providecommand \EOS [0]{\spacefactor3000\relax}%
\providecommand \BibitemShut  [1]{\csname bibitem#1\endcsname}%
\let\auto@bib@innerbib\@empty
\bibitem [{\citenamefont {Azzouz}(2006)}]{PhysRevB.74.174422}%
  \BibitemOpen
  \bibfield  {author} {\bibinfo {author} {\bibfnamefont {Mohamed}\ \bibnamefont
  {Azzouz}},\ }\bibfield  {title} {\enquote {\bibinfo {title} {Field-induced
  quantum criticality in low-dimensional heisenberg spin systems},}\ }\href
  {\doibase 10.1103/PhysRevB.74.174422} {\bibfield  {journal} {\bibinfo
  {journal} {Phys. Rev. B}\ }\textbf {\bibinfo {volume} {74}},\ \bibinfo
  {pages} {174422} (\bibinfo {year} {2006})}\BibitemShut {NoStop}%
\bibitem [{\citenamefont {Zhang}\ and\ \citenamefont
  {Song}(2015)}]{PhysRevLett.115.177204}%
  \BibitemOpen
  \bibfield  {author} {\bibinfo {author} {\bibfnamefont {G.}~\bibnamefont
  {Zhang}}\ and\ \bibinfo {author} {\bibfnamefont {Z.}~\bibnamefont {Song}},\
  }\bibfield  {title} {\enquote {\bibinfo {title} {Topological characterization
  of extended quantum ising models},}\ }\href {\doibase
  10.1103/PhysRevLett.115.177204} {\bibfield  {journal} {\bibinfo  {journal}
  {Phys. Rev. Lett.}\ }\textbf {\bibinfo {volume} {115}},\ \bibinfo {pages}
  {177204} (\bibinfo {year} {2015})}\BibitemShut {NoStop}%
\bibitem [{\citenamefont {Miao}\ \emph {et~al.}(2017)\citenamefont {Miao},
  \citenamefont {Jin}, \citenamefont {Zhang},\ and\ \citenamefont
  {Zhou}}]{PhysRevLett.118.267701}%
  \BibitemOpen
  \bibfield  {author} {\bibinfo {author} {\bibfnamefont {Jian-Jian}\
  \bibnamefont {Miao}}, \bibinfo {author} {\bibfnamefont {Hui-Ke}\ \bibnamefont
  {Jin}}, \bibinfo {author} {\bibfnamefont {Fu-Chun}\ \bibnamefont {Zhang}}, \
  and\ \bibinfo {author} {\bibfnamefont {Yi}~\bibnamefont {Zhou}},\ }\bibfield
  {title} {\enquote {\bibinfo {title} {Exact solution for the interacting
  kitaev chain at the symmetric point},}\ }\href {\doibase
  10.1103/PhysRevLett.118.267701} {\bibfield  {journal} {\bibinfo  {journal}
  {Phys. Rev. Lett.}\ }\textbf {\bibinfo {volume} {118}},\ \bibinfo {pages}
  {267701} (\bibinfo {year} {2017})}\BibitemShut {NoStop}%
\bibitem [{\citenamefont {Roy}\ \emph {et~al.}(2019)\citenamefont {Roy},
  \citenamefont {Chanda}, \citenamefont {Das}, \citenamefont {Sadhukhan},
  \citenamefont {Sen(De)},\ and\ \citenamefont {Sen}}]{PhysRevB.99.064422}%
  \BibitemOpen
  \bibfield  {author} {\bibinfo {author} {\bibfnamefont {Saptarshi}\
  \bibnamefont {Roy}}, \bibinfo {author} {\bibfnamefont {Titas}\ \bibnamefont
  {Chanda}}, \bibinfo {author} {\bibfnamefont {Tamoghna}\ \bibnamefont {Das}},
  \bibinfo {author} {\bibfnamefont {Debasis}\ \bibnamefont {Sadhukhan}},
  \bibinfo {author} {\bibfnamefont {Aditi}\ \bibnamefont {Sen(De)}}, \ and\
  \bibinfo {author} {\bibfnamefont {Ujjwal}\ \bibnamefont {Sen}},\ }\bibfield
  {title} {\enquote {\bibinfo {title} {Phase boundaries in an alternating-field
  quantum xy model with dzyaloshinskii-moriya interaction: Sustainable
  entanglement in dynamics},}\ }\href {\doibase 10.1103/PhysRevB.99.064422}
  {\bibfield  {journal} {\bibinfo  {journal} {Phys. Rev. B}\ }\textbf {\bibinfo
  {volume} {99}},\ \bibinfo {pages} {064422} (\bibinfo {year}
  {2019})}\BibitemShut {NoStop}%
\bibitem [{\citenamefont {Liu}\ \emph {et~al.}(2020)\citenamefont {Liu},
  \citenamefont {Yi}, \citenamefont {Sun}, \citenamefont {Dong},\ and\
  \citenamefont {You}}]{PhysRevE.102.032127}%
  \BibitemOpen
  \bibfield  {author} {\bibinfo {author} {\bibfnamefont {Zi-An}\ \bibnamefont
  {Liu}}, \bibinfo {author} {\bibfnamefont {Tian-Cheng}\ \bibnamefont {Yi}},
  \bibinfo {author} {\bibfnamefont {Jin-Hua}\ \bibnamefont {Sun}}, \bibinfo
  {author} {\bibfnamefont {Yu-Li}\ \bibnamefont {Dong}}, \ and\ \bibinfo
  {author} {\bibfnamefont {Wen-Long}\ \bibnamefont {You}},\ }\bibfield  {title}
  {\enquote {\bibinfo {title} {Lifshitz phase transitions in a one-dimensional
  gamma model},}\ }\href {\doibase 10.1103/PhysRevE.102.032127} {\bibfield
  {journal} {\bibinfo  {journal} {Phys. Rev. E}\ }\textbf {\bibinfo {volume}
  {102}},\ \bibinfo {pages} {032127} (\bibinfo {year} {2020})}\BibitemShut
  {NoStop}%
\bibitem [{\citenamefont {Luo}(2022)}]{PhysRevB.105.L060401}%
  \BibitemOpen
  \bibfield  {author} {\bibinfo {author} {\bibfnamefont {Qiang}\ \bibnamefont
  {Luo}},\ }\bibfield  {title} {\enquote {\bibinfo {title} {Analytical results
  for the unusual gr\"uneisen ratio in the quantum ising model with
  dzyaloshinskii-moriya interaction},}\ }\href {\doibase
  10.1103/PhysRevB.105.L060401} {\bibfield  {journal} {\bibinfo  {journal}
  {Phys. Rev. B}\ }\textbf {\bibinfo {volume} {105}},\ \bibinfo {pages}
  {L060401} (\bibinfo {year} {2022})}\BibitemShut {NoStop}%
\bibitem [{\citenamefont {Molignini}\ \emph {et~al.}(2017)\citenamefont
  {Molignini}, \citenamefont {van Nieuwenburg},\ and\ \citenamefont
  {Chitra}}]{Sensing_Floquet}%
  \BibitemOpen
  \bibfield  {author} {\bibinfo {author} {\bibfnamefont {Paolo}\ \bibnamefont
  {Molignini}}, \bibinfo {author} {\bibfnamefont {Evert}\ \bibnamefont {van
  Nieuwenburg}}, \ and\ \bibinfo {author} {\bibfnamefont {R.}~\bibnamefont
  {Chitra}},\ }\bibfield  {title} {\enquote {\bibinfo {title} {Sensing
  floquet-majorana fermions via heat transfer},}\ }\href {\doibase
  10.1103/PhysRevB.96.125144} {\bibfield  {journal} {\bibinfo  {journal} {Phys.
  Rev. B}\ }\textbf {\bibinfo {volume} {96}},\ \bibinfo {pages} {125144}
  (\bibinfo {year} {2017})}\BibitemShut {NoStop}%
\bibitem [{\citenamefont {Benenti}\ \emph {et~al.}(2009)\citenamefont
  {Benenti}, \citenamefont {Casati}, \citenamefont {Prosen}, \citenamefont
  {Rossini},\ and\ \citenamefont {\ifmmode \check{Z}\else
  \v{Z}\fi{}nidari\ifmmode~\check{c}\else
  \v{c}\fi{}}}]{Charge_and_spin_transport}%
  \BibitemOpen
  \bibfield  {author} {\bibinfo {author} {\bibfnamefont {Giuliano}\
  \bibnamefont {Benenti}}, \bibinfo {author} {\bibfnamefont {Giulio}\
  \bibnamefont {Casati}}, \bibinfo {author} {\bibfnamefont {Toma\ifmmode
  \check{z}\else~\v{z}\fi{}}\ \bibnamefont {Prosen}}, \bibinfo {author}
  {\bibfnamefont {Davide}\ \bibnamefont {Rossini}}, \ and\ \bibinfo {author}
  {\bibfnamefont {Marko}\ \bibnamefont {\ifmmode \check{Z}\else
  \v{Z}\fi{}nidari\ifmmode~\check{c}\else \v{c}\fi{}}},\ }\bibfield  {title}
  {\enquote {\bibinfo {title} {Charge and spin transport in strongly correlated
  one-dimensional quantum systems driven far from equilibrium},}\ }\href
  {\doibase 10.1103/PhysRevB.80.035110} {\bibfield  {journal} {\bibinfo
  {journal} {Phys. Rev. B}\ }\textbf {\bibinfo {volume} {80}},\ \bibinfo
  {pages} {035110} (\bibinfo {year} {2009})}\BibitemShut {NoStop}%
\bibitem [{\citenamefont {Bandyopadhyay}\ \emph {et~al.}(2021)\citenamefont
  {Bandyopadhyay}, \citenamefont {Bhattacharjee},\ and\ \citenamefont
  {Sen}}]{driven_topology}%
  \BibitemOpen
  \bibfield  {author} {\bibinfo {author} {\bibfnamefont {Souvik}\ \bibnamefont
  {Bandyopadhyay}}, \bibinfo {author} {\bibfnamefont {Sourav}\ \bibnamefont
  {Bhattacharjee}}, \ and\ \bibinfo {author} {\bibfnamefont {Diptiman}\
  \bibnamefont {Sen}},\ }\bibfield  {title} {\enquote {\bibinfo {title} {Driven
  quantum many-body systems and out-of-equilibrium topology},}\ }\href
  {\doibase 10.1088/1361-648X/ac1151} {\bibfield  {journal} {\bibinfo
  {journal} {Journal of Physics: Condensed Matter}\ }\textbf {\bibinfo {volume}
  {33}},\ \bibinfo {pages} {393001} (\bibinfo {year} {2021})}\BibitemShut
  {NoStop}%
\bibitem [{\citenamefont {Mondal}\ \emph {et~al.}(2022)\citenamefont {Mondal},
  \citenamefont {Sen},\ and\ \citenamefont
  {Dutta}}]{edge_state_entropy_Mondal_2023}%
  \BibitemOpen
  \bibfield  {author} {\bibinfo {author} {\bibfnamefont {Saikat}\ \bibnamefont
  {Mondal}}, \bibinfo {author} {\bibfnamefont {Diptiman}\ \bibnamefont {Sen}},
  \ and\ \bibinfo {author} {\bibfnamefont {Amit}\ \bibnamefont {Dutta}},\
  }\bibfield  {title} {\enquote {\bibinfo {title} {Disconnected entanglement
  entropy as a marker of edge modes in a periodically driven kitaev chain},}\
  }\href {\doibase 10.1088/1361-648X/aca7f7} {\bibfield  {journal} {\bibinfo
  {journal} {Journal of Physics: Condensed Matter}\ }\textbf {\bibinfo {volume}
  {35}},\ \bibinfo {pages} {085601} (\bibinfo {year} {2022})}\BibitemShut
  {NoStop}%
\bibitem [{\citenamefont {van Caspel}\ \emph {et~al.}(2019)\citenamefont {van
  Caspel}, \citenamefont {Arze},\ and\ \citenamefont
  {Castillo}}]{signature_dynamic_driven_kitaev}%
  \BibitemOpen
  \bibfield  {author} {\bibinfo {author} {\bibfnamefont {Moos}\ \bibnamefont
  {van Caspel}}, \bibinfo {author} {\bibfnamefont {Sergio Enrique~Tapias}\
  \bibnamefont {Arze}}, \ and\ \bibinfo {author} {\bibfnamefont {Isaac~Pérez}\
  \bibnamefont {Castillo}},\ }\bibfield  {title} {\enquote {\bibinfo {title}
  {{Dynamical signatures of topological order in the driven-dissipative Kitaev
  chain}},}\ }\href {\doibase 10.21468/SciPostPhys.6.2.026} {\bibfield
  {journal} {\bibinfo  {journal} {SciPost Phys.}\ }\textbf {\bibinfo {volume}
  {6}},\ \bibinfo {pages} {026} (\bibinfo {year} {2019})}\BibitemShut {NoStop}%
\bibitem [{\citenamefont {Rivas}\ and\ \citenamefont
  {Martin-Delgado}(2017)}]{nature_main_ref}%
  \BibitemOpen
  \bibfield  {author} {\bibinfo {author} {\bibfnamefont {{\'A}ngel}\
  \bibnamefont {Rivas}}\ and\ \bibinfo {author} {\bibfnamefont {Miguel~A.}\
  \bibnamefont {Martin-Delgado}},\ }\bibfield  {title} {\enquote {\bibinfo
  {title} {Topological heat transport and symmetry-protected boson currents},}\
  }\href {\doibase 10.1038/s41598-017-06722-x} {\bibfield  {journal} {\bibinfo
  {journal} {Scientific Reports}\ }\textbf {\bibinfo {volume} {7}},\ \bibinfo
  {pages} {6350} (\bibinfo {year} {2017})}\BibitemShut {NoStop}%
\bibitem [{\citenamefont {Kane}\ and\ \citenamefont
  {Moore}(2011)}]{Kane_topo_review}%
  \BibitemOpen
  \bibfield  {author} {\bibinfo {author} {\bibfnamefont {Charles}\ \bibnamefont
  {Kane}}\ and\ \bibinfo {author} {\bibfnamefont {Joel}\ \bibnamefont
  {Moore}},\ }\bibfield  {title} {\enquote {\bibinfo {title} {Topological
  insulators},}\ }\href {\doibase 10.1088/2058-7058/24/02/36} {\bibfield
  {journal} {\bibinfo  {journal} {Physics World}\ }\textbf {\bibinfo {volume}
  {24}},\ \bibinfo {pages} {32} (\bibinfo {year} {2011})}\BibitemShut {NoStop}%
\bibitem [{\citenamefont {Asb{\'{o}}th}\ \emph {et~al.}(2016)\citenamefont
  {Asb{\'{o}}th}, \citenamefont {Oroszl{\'{a}}ny},\ and\ \citenamefont
  {P{\'{a}}lyi}}]{short_course_topo}%
  \BibitemOpen
  \bibfield  {author} {\bibinfo {author} {\bibfnamefont {J{\'{a} }nos~K.}\
  \bibnamefont {Asb{\'{o}}th}}, \bibinfo {author} {\bibfnamefont
  {L{\'{a}}szl{\'{o}}}\ \bibnamefont {Oroszl{\'{a}}ny}}, \ and\ \bibinfo
  {author} {\bibfnamefont {Andr{\'{a}}s}\ \bibnamefont {P{\'{a}}lyi}},\ }\href
  {\doibase 10.1007/978-3-319-25607-8} {\emph {\bibinfo {title} {A Short Course
  on Topological Insulators}}}\ (\bibinfo  {publisher} {Springer International
  Publishing},\ \bibinfo {year} {2016})\BibitemShut {NoStop}%
\bibitem [{\citenamefont {Viyuela}\ \emph {et~al.}(2014)\citenamefont
  {Viyuela}, \citenamefont {Rivas},\ and\ \citenamefont
  {Martin-Delgado}}]{Ulham_ref}%
  \BibitemOpen
  \bibfield  {author} {\bibinfo {author} {\bibfnamefont {O.}~\bibnamefont
  {Viyuela}}, \bibinfo {author} {\bibfnamefont {A.}~\bibnamefont {Rivas}}, \
  and\ \bibinfo {author} {\bibfnamefont {M.{\hspace{0.167em} }A.}\ \bibnamefont
  {Martin-Delgado}},\ }\bibfield  {title} {\enquote {\bibinfo {title} {Uhlmann
  phase as a topological measure for one-dimensional fermion systems},}\ }\href
  {https://doi.org/10.1103%2Fphysrevlett.112.130401} {\bibfield  {journal}
  {\bibinfo  {journal} {Physical Review Letters}\ }\textbf {\bibinfo {volume}
  {112}},\ \bibinfo {pages} {130401} (\bibinfo {year} {2014})}\BibitemShut
  {NoStop}%
\bibitem [{\citenamefont {Kempkes}\ \emph {et~al.}(2016)\citenamefont
  {Kempkes}, \citenamefont {Quelle},\ and\ \citenamefont
  {Smith}}]{hill_kempkes_ref}%
  \BibitemOpen
  \bibfield  {author} {\bibinfo {author} {\bibfnamefont {S.~N.}\ \bibnamefont
  {Kempkes}}, \bibinfo {author} {\bibfnamefont {A.}~\bibnamefont {Quelle}}, \
  and\ \bibinfo {author} {\bibfnamefont {C.~Morais}\ \bibnamefont {Smith}},\
  }\bibfield  {title} {\enquote {\bibinfo {title} {Universalities of
  thermodynamic signatures in topological phases},}\ }\href {\doibase
  10.1038/srep38530} {\bibfield  {journal} {\bibinfo  {journal} {Scientific
  Reports}\ }\textbf {\bibinfo {volume} {6}},\ \bibinfo {pages} {38530}
  (\bibinfo {year} {2016})}\BibitemShut {NoStop}%
\bibitem [{\citenamefont {Meier}\ \emph
  {et~al.}(2016{\natexlab{a}})\citenamefont {Meier}, \citenamefont {An},\ and\
  \citenamefont {Gadway}}]{exp1}%
  \BibitemOpen
  \bibfield  {author} {\bibinfo {author} {\bibfnamefont {Eric~J.}\ \bibnamefont
  {Meier}}, \bibinfo {author} {\bibfnamefont {Fangzhao~Alex}\ \bibnamefont
  {An}}, \ and\ \bibinfo {author} {\bibfnamefont {Bryce}\ \bibnamefont
  {Gadway}},\ }\bibfield  {title} {\enquote {\bibinfo {title} {Observation of
  the topological soliton state in the su--schrieffer--heeger model},}\ }\href
  {\doibase 10.1038/ncomms13986} {\bibfield  {journal} {\bibinfo  {journal}
  {Nature Communications}\ }\textbf {\bibinfo {volume} {7}},\ \bibinfo {pages}
  {13986} (\bibinfo {year} {2016}{\natexlab{a}})}\BibitemShut {NoStop}%
\bibitem [{\citenamefont {Drost}\ \emph {et~al.}(2017)\citenamefont {Drost},
  \citenamefont {Ojanen}, \citenamefont {Harju},\ and\ \citenamefont
  {Liljeroth}}]{exp2}%
  \BibitemOpen
  \bibfield  {author} {\bibinfo {author} {\bibfnamefont {Robert}\ \bibnamefont
  {Drost}}, \bibinfo {author} {\bibfnamefont {Teemu}\ \bibnamefont {Ojanen}},
  \bibinfo {author} {\bibfnamefont {Ari}\ \bibnamefont {Harju}}, \ and\
  \bibinfo {author} {\bibfnamefont {Peter}\ \bibnamefont {Liljeroth}},\
  }\bibfield  {title} {\enquote {\bibinfo {title} {Topological states in
  engineered atomic lattices},}\ }\href {\doibase 10.1038/nphys4080} {\bibfield
   {journal} {\bibinfo  {journal} {Nature Physics}\ }\textbf {\bibinfo {volume}
  {13}},\ \bibinfo {pages} {668--671} (\bibinfo {year} {2017})}\BibitemShut
  {NoStop}%
\bibitem [{\citenamefont {Kanungo}\ \emph {et~al.}(2022)\citenamefont
  {Kanungo}, \citenamefont {Whalen}, \citenamefont {Lu}, \citenamefont {Yuan},
  \citenamefont {Dasgupta}, \citenamefont {Dunning}, \citenamefont {Hazzard},\
  and\ \citenamefont {Killian}}]{exp3}%
  \BibitemOpen
  \bibfield  {author} {\bibinfo {author} {\bibfnamefont {S.~K.}\ \bibnamefont
  {Kanungo}}, \bibinfo {author} {\bibfnamefont {J.~D.}\ \bibnamefont {Whalen}},
  \bibinfo {author} {\bibfnamefont {Y.}~\bibnamefont {Lu}}, \bibinfo {author}
  {\bibfnamefont {M.}~\bibnamefont {Yuan}}, \bibinfo {author} {\bibfnamefont
  {S.}~\bibnamefont {Dasgupta}}, \bibinfo {author} {\bibfnamefont {F.~B.}\
  \bibnamefont {Dunning}}, \bibinfo {author} {\bibfnamefont {K.~R.~A.}\
  \bibnamefont {Hazzard}}, \ and\ \bibinfo {author} {\bibfnamefont {T.~C.}\
  \bibnamefont {Killian}},\ }\bibfield  {title} {\enquote {\bibinfo {title}
  {Realizing topological edge states with rydberg-atom synthetic dimensions},}\
  }\href {\doibase 10.1038/s41467-022-28550-y} {\bibfield  {journal} {\bibinfo
  {journal} {Nature Communications}\ }\textbf {\bibinfo {volume} {13}},\
  \bibinfo {pages} {972} (\bibinfo {year} {2022})}\BibitemShut {NoStop}%
\bibitem [{\citenamefont {Meier}\ \emph
  {et~al.}(2016{\natexlab{b}})\citenamefont {Meier}, \citenamefont {An},\ and\
  \citenamefont {Gadway}}]{exp4}%
  \BibitemOpen
  \bibfield  {author} {\bibinfo {author} {\bibfnamefont {Eric~J.}\ \bibnamefont
  {Meier}}, \bibinfo {author} {\bibfnamefont {Fangzhao~Alex}\ \bibnamefont
  {An}}, \ and\ \bibinfo {author} {\bibfnamefont {Bryce}\ \bibnamefont
  {Gadway}},\ }\bibfield  {title} {\enquote {\bibinfo {title} {Observation of
  the topological soliton state in the su--schrieffer--heeger model},}\ }\href
  {\doibase 10.1038/ncomms13986} {\bibfield  {journal} {\bibinfo  {journal}
  {Nature Communications}\ }\textbf {\bibinfo {volume} {7}},\ \bibinfo {pages}
  {13986} (\bibinfo {year} {2016}{\natexlab{b}})}\BibitemShut {NoStop}%
\bibitem [{\citenamefont {Gr{\"o}ning}\ \emph {et~al.}(2018)\citenamefont
  {Gr{\"o}ning}, \citenamefont {Wang}, \citenamefont {Yao}, \citenamefont
  {Pignedoli}, \citenamefont {Borin~Barin}, \citenamefont {Daniels},
  \citenamefont {Cupo}, \citenamefont {Meunier}, \citenamefont {Feng},
  \citenamefont {Narita}, \citenamefont {M{\"u}llen}, \citenamefont
  {Ruffieux},\ and\ \citenamefont {Fasel}}]{exp5}%
  \BibitemOpen
  \bibfield  {author} {\bibinfo {author} {\bibfnamefont {Oliver}\ \bibnamefont
  {Gr{\"o}ning}}, \bibinfo {author} {\bibfnamefont {Shiyong}\ \bibnamefont
  {Wang}}, \bibinfo {author} {\bibfnamefont {Xuelin}\ \bibnamefont {Yao}},
  \bibinfo {author} {\bibfnamefont {Carlo~A}\ \bibnamefont {Pignedoli}},
  \bibinfo {author} {\bibfnamefont {Gabriela}\ \bibnamefont {Borin~Barin}},
  \bibinfo {author} {\bibfnamefont {Colin}\ \bibnamefont {Daniels}}, \bibinfo
  {author} {\bibfnamefont {Andrew}\ \bibnamefont {Cupo}}, \bibinfo {author}
  {\bibfnamefont {Vincent}\ \bibnamefont {Meunier}}, \bibinfo {author}
  {\bibfnamefont {Xinliang}\ \bibnamefont {Feng}}, \bibinfo {author}
  {\bibfnamefont {Akimitsu}\ \bibnamefont {Narita}}, \bibinfo {author}
  {\bibfnamefont {Klaus}\ \bibnamefont {M{\"u}llen}}, \bibinfo {author}
  {\bibfnamefont {Pascal}\ \bibnamefont {Ruffieux}}, \ and\ \bibinfo {author}
  {\bibfnamefont {Roman"}\ \bibnamefont {Fasel}},\ }\bibfield  {title}
  {\enquote {\bibinfo {title} {Engineering of robust topological quantum phases
  in graphene nanoribbons},}\ }\href {\doibase 10.1038/s41586-018-0375-9}
  {\bibfield  {journal} {\bibinfo  {journal} {Nature}\ }\textbf {\bibinfo
  {volume} {560}},\ \bibinfo {pages} {209--213} (\bibinfo {year}
  {2018})}\BibitemShut {NoStop}%
\bibitem [{\citenamefont {Tanaka}\ \emph {et~al.}(2022)\citenamefont {Tanaka},
  \citenamefont {Lu},\ and\ \citenamefont {Nagaosa}}]{topo_diode1}%
  \BibitemOpen
  \bibfield  {author} {\bibinfo {author} {\bibfnamefont {Yukio}\ \bibnamefont
  {Tanaka}}, \bibinfo {author} {\bibfnamefont {Bo}~\bibnamefont {Lu}}, \ and\
  \bibinfo {author} {\bibfnamefont {Naoto}\ \bibnamefont {Nagaosa}},\
  }\bibfield  {title} {\enquote {\bibinfo {title} {Theory of giant diode effect
  in $d$-wave superconductor junctions on the surface of a topological
  insulator},}\ }\href {\doibase 10.1103/PhysRevB.106.214524} {\bibfield
  {journal} {\bibinfo  {journal} {Phys. Rev. B}\ }\textbf {\bibinfo {volume}
  {106}},\ \bibinfo {pages} {214524} (\bibinfo {year} {2022})}\BibitemShut
  {NoStop}%
\bibitem [{\citenamefont {Fluckey}\ \emph {et~al.}(2022)\citenamefont
  {Fluckey}, \citenamefont {Tiwari}, \citenamefont {Hinkle},\ and\
  \citenamefont {Vandenberghe}}]{topo_diode2}%
  \BibitemOpen
  \bibfield  {author} {\bibinfo {author} {\bibfnamefont {Stephen~P.}\
  \bibnamefont {Fluckey}}, \bibinfo {author} {\bibfnamefont {Sabyasachi}\
  \bibnamefont {Tiwari}}, \bibinfo {author} {\bibfnamefont {Christopher~L.}\
  \bibnamefont {Hinkle}}, \ and\ \bibinfo {author} {\bibfnamefont {William~G.}\
  \bibnamefont {Vandenberghe}},\ }\bibfield  {title} {\enquote {\bibinfo
  {title} {Three-dimensional-topological-insulator tunnel diodes},}\ }\href
  {\doibase 10.1103/PhysRevApplied.18.064037} {\bibfield  {journal} {\bibinfo
  {journal} {Phys. Rev. Appl.}\ }\textbf {\bibinfo {volume} {18}},\ \bibinfo
  {pages} {064037} (\bibinfo {year} {2022})}\BibitemShut {NoStop}%
\bibitem [{\citenamefont {Karabassov}\ \emph {et~al.}(2023)\citenamefont
  {Karabassov}, \citenamefont {Amirov}, \citenamefont {Bobkova}, \citenamefont
  {Golubov}, \citenamefont {Kazakova},\ and\ \citenamefont
  {Vasenko}}]{topo_diode3}%
  \BibitemOpen
  \bibfield  {author} {\bibinfo {author} {\bibfnamefont {Tairzhan}\
  \bibnamefont {Karabassov}}, \bibinfo {author} {\bibfnamefont {Emir~S.}\
  \bibnamefont {Amirov}}, \bibinfo {author} {\bibfnamefont {Irina~V.}\
  \bibnamefont {Bobkova}}, \bibinfo {author} {\bibfnamefont {Alexander~A.}\
  \bibnamefont {Golubov}}, \bibinfo {author} {\bibfnamefont {Elena~A.}\
  \bibnamefont {Kazakova}}, \ and\ \bibinfo {author} {\bibfnamefont
  {Andrey~S.}\ \bibnamefont {Vasenko}},\ }\bibfield  {title} {\enquote
  {\bibinfo {title} {Superconducting diode effect in topological hybrid
  structures},}\ }\href {https://www.mdpi.com/2410-3896/8/2/36} {\bibfield
  {journal} {\bibinfo  {journal} {Condensed Matter}\ }\textbf {\bibinfo
  {volume} {8}} (\bibinfo {year} {2023})}\BibitemShut {NoStop}%
\bibitem [{\citenamefont {Li}\ \emph {et~al.}(2017)\citenamefont {Li},
  \citenamefont {Zhou}, \citenamefont {Li}, \citenamefont {Che}, \citenamefont
  {Yao}, \citenamefont {McHale}, \citenamefont {Chaudhury},\ and\ \citenamefont
  {Wang}}]{topo_diode_liquid}%
  \BibitemOpen
  \bibfield  {author} {\bibinfo {author} {\bibfnamefont {Jiaqian}\ \bibnamefont
  {Li}}, \bibinfo {author} {\bibfnamefont {Xiaofeng}\ \bibnamefont {Zhou}},
  \bibinfo {author} {\bibfnamefont {Jing}\ \bibnamefont {Li}}, \bibinfo
  {author} {\bibfnamefont {Lufeng}\ \bibnamefont {Che}}, \bibinfo {author}
  {\bibfnamefont {Jun}\ \bibnamefont {Yao}}, \bibinfo {author} {\bibfnamefont
  {Glen}\ \bibnamefont {McHale}}, \bibinfo {author} {\bibfnamefont {Manoj~K.}\
  \bibnamefont {Chaudhury}}, \ and\ \bibinfo {author} {\bibfnamefont {Zuankai}\
  \bibnamefont {Wang}},\ }\bibfield  {title} {\enquote {\bibinfo {title}
  {Topological liquid diode},}\ }\href {\doibase 10.1126/sciadv.aao3530}
  {\bibfield  {journal} {\bibinfo  {journal} {Science Advances}\ }\textbf
  {\bibinfo {volume} {3}},\ \bibinfo {pages} {eaao3530} (\bibinfo {year}
  {2017})}\BibitemShut {NoStop}%
\bibitem [{\citenamefont {Quelle}\ \emph {et~al.}(2016)\citenamefont {Quelle},
  \citenamefont {Cobanera},\ and\ \citenamefont
  {Smith}}]{Quelle_PhysRevB.94.075133}%
  \BibitemOpen
  \bibfield  {author} {\bibinfo {author} {\bibfnamefont {A.}~\bibnamefont
  {Quelle}}, \bibinfo {author} {\bibfnamefont {E.}~\bibnamefont {Cobanera}}, \
  and\ \bibinfo {author} {\bibfnamefont {C.~Morais}\ \bibnamefont {Smith}},\
  }\bibfield  {title} {\enquote {\bibinfo {title} {Thermodynamic signatures of
  edge states in topological insulators},}\ }\href {\doibase
  10.1103/PhysRevB.94.075133} {\bibfield  {journal} {\bibinfo  {journal} {Phys.
  Rev. B}\ }\textbf {\bibinfo {volume} {94}},\ \bibinfo {pages} {075133}
  (\bibinfo {year} {2016})}\BibitemShut {NoStop}%
\bibitem [{\citenamefont {He}\ and\ \citenamefont {Chien}(2023)}]{similar_2}%
  \BibitemOpen
  \bibfield  {author} {\bibinfo {author} {\bibfnamefont {Yan}\ \bibnamefont
  {He}}\ and\ \bibinfo {author} {\bibfnamefont {Chih-Chun}\ \bibnamefont
  {Chien}},\ }\bibfield  {title} {\enquote {\bibinfo {title} {Particle and
  thermal transport through one dimensional topological systems via lindblad
  formalism},}\ }\href {\doibase
  https://doi.org/10.1016/j.physleta.2023.128826} {\bibfield  {journal}
  {\bibinfo  {journal} {Physics Letters A}\ }\textbf {\bibinfo {volume}
  {473}},\ \bibinfo {pages} {128826} (\bibinfo {year} {2023})}\BibitemShut
  {NoStop}%
\bibitem [{\citenamefont {Nava}\ \emph {et~al.}(2023)\citenamefont {Nava},
  \citenamefont {Campagnano}, \citenamefont {Sodano},\ and\ \citenamefont
  {Giuliano}}]{similar_ssh}%
  \BibitemOpen
  \bibfield  {author} {\bibinfo {author} {\bibfnamefont {Andrea}\ \bibnamefont
  {Nava}}, \bibinfo {author} {\bibfnamefont {Gabriele}\ \bibnamefont
  {Campagnano}}, \bibinfo {author} {\bibfnamefont {Pasquale}\ \bibnamefont
  {Sodano}}, \ and\ \bibinfo {author} {\bibfnamefont {Domenico}\ \bibnamefont
  {Giuliano}},\ }\bibfield  {title} {\enquote {\bibinfo {title} {Lindblad
  master equation approach to the topological phase transition in the
  disordered su-schrieffer-heeger model},}\ }\href {\doibase
  10.1103/PhysRevB.107.035113} {\bibfield  {journal} {\bibinfo  {journal}
  {Phys. Rev. B}\ }\textbf {\bibinfo {volume} {107}},\ \bibinfo {pages}
  {035113} (\bibinfo {year} {2023})}\BibitemShut {NoStop}%
\bibitem [{\citenamefont {Levy}\ and\ \citenamefont
  {Kosloff}(2014)}]{Levy_2014}%
  \BibitemOpen
  \bibfield  {author} {\bibinfo {author} {\bibfnamefont {Amikam}\ \bibnamefont
  {Levy}}\ and\ \bibinfo {author} {\bibfnamefont {Ronnie}\ \bibnamefont
  {Kosloff}},\ }\bibfield  {title} {\enquote {\bibinfo {title} {The local
  approach to quantum transport may violate the second law of
  thermodynamics},}\ }\href {\doibase 10.1209/0295-5075/107/20004} {\bibfield
  {journal} {\bibinfo  {journal} {Europhys. Lett.}\ }\textbf {\bibinfo {volume}
  {107}},\ \bibinfo {pages} {20004} (\bibinfo {year} {2014})}\BibitemShut
  {NoStop}%
\bibitem [{\citenamefont {Naseem}\ \emph {et~al.}(2018)\citenamefont {Naseem},
  \citenamefont {Xuereb},\ and\ \citenamefont {M\"ustecapl\ifmmode \imath \else
  \i \fi{}o\ifmmode~\breve{g}\else \u{g}\fi{}lu}}]{PhysRevA.98.052123}%
  \BibitemOpen
  \bibfield  {author} {\bibinfo {author} {\bibfnamefont {M.~Tahir}\
  \bibnamefont {Naseem}}, \bibinfo {author} {\bibfnamefont {Andr\'e}\
  \bibnamefont {Xuereb}}, \ and\ \bibinfo {author} {\bibfnamefont
  {\"Ozg\"ur~E.}\ \bibnamefont {M\"ustecapl\ifmmode \imath \else \i
  \fi{}o\ifmmode~\breve{g}\else \u{g}\fi{}lu}},\ }\bibfield  {title} {\enquote
  {\bibinfo {title} {Thermodynamic consistency of the optomechanical master
  equation},}\ }\href {\doibase 10.1103/PhysRevA.98.052123} {\bibfield
  {journal} {\bibinfo  {journal} {Phys. Rev. A}\ }\textbf {\bibinfo {volume}
  {98}},\ \bibinfo {pages} {052123} (\bibinfo {year} {2018})}\BibitemShut
  {NoStop}%
\bibitem [{\citenamefont {Landi}\ \emph {et~al.}(2014)\citenamefont {Landi},
  \citenamefont {Novais}, \citenamefont {de~Oliveira},\ and\ \citenamefont
  {Karevski}}]{diode1}%
  \BibitemOpen
  \bibfield  {author} {\bibinfo {author} {\bibfnamefont {Gabriel~T.}\
  \bibnamefont {Landi}}, \bibinfo {author} {\bibfnamefont {E.}~\bibnamefont
  {Novais}}, \bibinfo {author} {\bibfnamefont {M\'ario~J.}\ \bibnamefont
  {de~Oliveira}}, \ and\ \bibinfo {author} {\bibfnamefont {Dragi}\ \bibnamefont
  {Karevski}},\ }\bibfield  {title} {\enquote {\bibinfo {title} {Flux
  rectification in the quantum $xxz$ chain},}\ }\href {\doibase
  10.1103/PhysRevE.90.042142} {\bibfield  {journal} {\bibinfo  {journal} {Phys.
  Rev. E}\ }\textbf {\bibinfo {volume} {90}},\ \bibinfo {pages} {042142}
  (\bibinfo {year} {2014})}\BibitemShut {NoStop}%
\bibitem [{\citenamefont {Jiang}\ \emph {et~al.}(2010)\citenamefont {Jiang},
  \citenamefont {Wang},\ and\ \citenamefont {Li}}]{diode2}%
  \BibitemOpen
  \bibfield  {author} {\bibinfo {author} {\bibfnamefont {J.~W.}\ \bibnamefont
  {Jiang}}, \bibinfo {author} {\bibfnamefont {J.~S.}\ \bibnamefont {Wang}}, \
  and\ \bibinfo {author} {\bibfnamefont {B.}~\bibnamefont {Li}},\ }\bibfield
  {title} {\enquote {\bibinfo {title} {Topology-induced thermal rectification
  in carbon nanodevice},}\ }\href {\doibase 10.1209/0295-5075/89/46005}
  {\bibfield  {journal} {\bibinfo  {journal} {Europhys. Lett.}\ }\textbf
  {\bibinfo {volume} {89}},\ \bibinfo {pages} {46005} (\bibinfo {year}
  {2010})}\BibitemShut {NoStop}%
\bibitem [{\citenamefont {Ordonez-Miranda}\ \emph {et~al.}(2017)\citenamefont
  {Ordonez-Miranda}, \citenamefont {Ezzahri},\ and\ \citenamefont
  {Joulain}}]{diode4}%
  \BibitemOpen
  \bibfield  {author} {\bibinfo {author} {\bibfnamefont {Jose}\ \bibnamefont
  {Ordonez-Miranda}}, \bibinfo {author} {\bibfnamefont {Youn\`es}\ \bibnamefont
  {Ezzahri}}, \ and\ \bibinfo {author} {\bibfnamefont {Karl}\ \bibnamefont
  {Joulain}},\ }\bibfield  {title} {\enquote {\bibinfo {title} {Quantum thermal
  diode based on two interacting spinlike systems under different
  excitations},}\ }\href {\doibase 10.1103/PhysRevE.95.022128} {\bibfield
  {journal} {\bibinfo  {journal} {Phys. Rev. E}\ }\textbf {\bibinfo {volume}
  {95}},\ \bibinfo {pages} {022128} (\bibinfo {year} {2017})}\BibitemShut
  {NoStop}%
\bibitem [{\citenamefont {Karg\ifmmode \imath \else~\i \fi{}}\ \emph
  {et~al.}(2019)\citenamefont {Karg\ifmmode \imath \else~\i \fi{}},
  \citenamefont {Naseem}, \citenamefont {Opatrn\'y}, \citenamefont
  {M\"ustecapl\ifmmode \imath \else \i \fi{}o\ifmmode~\breve{g}\else
  \u{g}\fi{}lu},\ and\ \citenamefont {Kurizki}}]{diode6}%
  \BibitemOpen
  \bibfield  {author} {\bibinfo {author} {\bibfnamefont {Cahit}\ \bibnamefont
  {Karg\ifmmode \imath \else~\i \fi{}}}, \bibinfo {author} {\bibfnamefont
  {M.~Tahir}\ \bibnamefont {Naseem}}, \bibinfo {author} {\bibfnamefont
  {Tom\'a\ifmmode \check{s}\else~\v{s}\fi{}}\ \bibnamefont {Opatrn\'y}},
  \bibinfo {author} {\bibfnamefont {\"Ozg\"ur~E.}\ \bibnamefont
  {M\"ustecapl\ifmmode \imath \else \i \fi{}o\ifmmode~\breve{g}\else
  \u{g}\fi{}lu}}, \ and\ \bibinfo {author} {\bibfnamefont {Gershon}\
  \bibnamefont {Kurizki}},\ }\bibfield  {title} {\enquote {\bibinfo {title}
  {Quantum optical two-atom thermal diode},}\ }\href {\doibase
  10.1103/PhysRevE.99.042121} {\bibfield  {journal} {\bibinfo  {journal} {Phys.
  Rev. E}\ }\textbf {\bibinfo {volume} {99}},\ \bibinfo {pages} {042121}
  (\bibinfo {year} {2019})}\BibitemShut {NoStop}%
\bibitem [{\citenamefont {Segal}\ and\ \citenamefont
  {Nitzan}(2005)}]{dvira_segal_diode}%
  \BibitemOpen
  \bibfield  {author} {\bibinfo {author} {\bibfnamefont {Dvira}\ \bibnamefont
  {Segal}}\ and\ \bibinfo {author} {\bibfnamefont {Abraham}\ \bibnamefont
  {Nitzan}},\ }\bibfield  {title} {\enquote {\bibinfo {title} {Spin-boson
  thermal rectifier},}\ }\href {\doibase 10.1103/PhysRevLett.94.034301}
  {\bibfield  {journal} {\bibinfo  {journal} {Phys. Rev. Lett.}\ }\textbf
  {\bibinfo {volume} {94}},\ \bibinfo {pages} {034301} (\bibinfo {year}
  {2005})}\BibitemShut {NoStop}%
\bibitem [{\citenamefont {Su}\ \emph {et~al.}(1979)\citenamefont {Su},
  \citenamefont {Schrieffer},\ and\ \citenamefont
  {Heeger}}]{ssh_original_paper}%
  \BibitemOpen
  \bibfield  {author} {\bibinfo {author} {\bibfnamefont {W.~P.}\ \bibnamefont
  {Su}}, \bibinfo {author} {\bibfnamefont {J.~R.}\ \bibnamefont {Schrieffer}},
  \ and\ \bibinfo {author} {\bibfnamefont {A.~J.}\ \bibnamefont {Heeger}},\
  }\bibfield  {title} {\enquote {\bibinfo {title} {Solitons in
  polyacetylene},}\ }\href {\doibase 10.1103/PhysRevLett.42.1698} {\bibfield
  {journal} {\bibinfo  {journal} {Phys. Rev. Lett.}\ }\textbf {\bibinfo
  {volume} {42}},\ \bibinfo {pages} {1698--1701} (\bibinfo {year}
  {1979})}\BibitemShut {NoStop}%
\bibitem [{\citenamefont {Sirker}\ \emph {et~al.}(2014)\citenamefont {Sirker},
  \citenamefont {Maiti}, \citenamefont {Konstantinidis},\ and\ \citenamefont
  {Sedlmayr}}]{Ssh_eigen}%
  \BibitemOpen
  \bibfield  {author} {\bibinfo {author} {\bibfnamefont {J}~\bibnamefont
  {Sirker}}, \bibinfo {author} {\bibfnamefont {M}~\bibnamefont {Maiti}},
  \bibinfo {author} {\bibfnamefont {N~P}\ \bibnamefont {Konstantinidis}}, \
  and\ \bibinfo {author} {\bibfnamefont {N}~\bibnamefont {Sedlmayr}},\
  }\bibfield  {title} {\enquote {\bibinfo {title} {Boundary fidelity and
  entanglement in the symmetry protected topological phase of the {SSH}
  model},}\ }\href {\doibase 10.1088/1742-5468/2014/10/p10032} {\bibfield
  {journal} {\bibinfo  {journal} {Journal of Statistical Mechanics: Theory and
  Experiment}\ }\textbf {\bibinfo {volume} {2014}},\ \bibinfo {pages} {P10032}
  (\bibinfo {year} {2014})}\BibitemShut {NoStop}%
\bibitem [{\citenamefont {Lacroix}\ \emph {et~al.}(2020)\citenamefont
  {Lacroix}, \citenamefont {Sargsyan}, \citenamefont {Adamian}, \citenamefont
  {Antonenko},\ and\ \citenamefont
  {Hovhannisyan}}]{System_Bath_PhysRevA.102.022209}%
  \BibitemOpen
  \bibfield  {author} {\bibinfo {author} {\bibfnamefont {Denis}\ \bibnamefont
  {Lacroix}}, \bibinfo {author} {\bibfnamefont {V.~V.}\ \bibnamefont
  {Sargsyan}}, \bibinfo {author} {\bibfnamefont {G.~G.}\ \bibnamefont
  {Adamian}}, \bibinfo {author} {\bibfnamefont {N.~V.}\ \bibnamefont
  {Antonenko}}, \ and\ \bibinfo {author} {\bibfnamefont {A.~A.}\ \bibnamefont
  {Hovhannisyan}},\ }\bibfield  {title} {\enquote {\bibinfo {title}
  {Non-markovian modeling of fermi-bose systems coupled to one or several
  fermi-bose thermal baths},}\ }\href {\doibase 10.1103/PhysRevA.102.022209}
  {\bibfield  {journal} {\bibinfo  {journal} {Phys. Rev. A}\ }\textbf {\bibinfo
  {volume} {102}},\ \bibinfo {pages} {022209} (\bibinfo {year}
  {2020})}\BibitemShut {NoStop}%
\bibitem [{\citenamefont {Sargsyan}\ \emph {et~al.}(2018)\citenamefont
  {Sargsyan}, \citenamefont {Hovhannisyan}, \citenamefont {Adamian},
  \citenamefont {Antonenko},\ and\ \citenamefont
  {Lacroix}}]{System_bath_SARGSYAN2018666}%
  \BibitemOpen
  \bibfield  {author} {\bibinfo {author} {\bibfnamefont {V.V.}\ \bibnamefont
  {Sargsyan}}, \bibinfo {author} {\bibfnamefont {A.A.}\ \bibnamefont
  {Hovhannisyan}}, \bibinfo {author} {\bibfnamefont {G.G.}\ \bibnamefont
  {Adamian}}, \bibinfo {author} {\bibfnamefont {N.V.}\ \bibnamefont
  {Antonenko}}, \ and\ \bibinfo {author} {\bibfnamefont {D.}~\bibnamefont
  {Lacroix}},\ }\bibfield  {title} {\enquote {\bibinfo {title} {Non-markovian
  dynamics of mixed fermionic–bosonic systems: Full coupling},}\ }\href
  {\doibase https://doi.org/10.1016/j.physa.2018.04.008} {\bibfield  {journal}
  {\bibinfo  {journal} {Physica A: Statistical Mechanics and its Applications}\
  }\textbf {\bibinfo {volume} {505}},\ \bibinfo {pages} {666--679} (\bibinfo
  {year} {2018})}\BibitemShut {NoStop}%
\bibitem [{\citenamefont {Hovhannisyan}\ \emph {et~al.}(2018)\citenamefont
  {Hovhannisyan}, \citenamefont {Sargsyan}, \citenamefont {Adamian},
  \citenamefont {Antonenko},\ and\ \citenamefont
  {Lacroix}}]{System_bath_PhysRevE.97.032134}%
  \BibitemOpen
  \bibfield  {author} {\bibinfo {author} {\bibfnamefont {A.~A.}\ \bibnamefont
  {Hovhannisyan}}, \bibinfo {author} {\bibfnamefont {V.~V.}\ \bibnamefont
  {Sargsyan}}, \bibinfo {author} {\bibfnamefont {G.~G.}\ \bibnamefont
  {Adamian}}, \bibinfo {author} {\bibfnamefont {N.~V.}\ \bibnamefont
  {Antonenko}}, \ and\ \bibinfo {author} {\bibfnamefont {D.}~\bibnamefont
  {Lacroix}},\ }\bibfield  {title} {\enquote {\bibinfo {title} {Non-markovian
  dynamics of fermionic and bosonic systems coupled to several heat baths},}\
  }\href {\doibase 10.1103/PhysRevE.97.032134} {\bibfield  {journal} {\bibinfo
  {journal} {Phys. Rev. E}\ }\textbf {\bibinfo {volume} {97}},\ \bibinfo
  {pages} {032134} (\bibinfo {year} {2018})}\BibitemShut {NoStop}%
\bibitem [{\citenamefont {Ikramov}(1999)}]{ssh_eigen_2}%
  \BibitemOpen
  \bibfield  {author} {\bibinfo {author} {\bibfnamefont {Khakim~D.}\
  \bibnamefont {Ikramov}},\ }\bibfield  {title} {\enquote {\bibinfo {title}
  {Shin's formulas for eigenpairs of symmetric tridiagonal 2-toeplitz
  matrices},}\ }\href {\doibase 10.1017/S0004972700032664} {\bibfield
  {journal} {\bibinfo  {journal} {Bulletin of the Australian Mathematical
  Society}\ }\textbf {\bibinfo {volume} {59}},\ \bibinfo {pages} {119–120}
  (\bibinfo {year} {1999})}\BibitemShut {NoStop}%
\bibitem [{\citenamefont {Breuer}\ and\ \citenamefont
  {Petruccione}(2002)}]{breuer2002}%
  \BibitemOpen
  \bibfield  {author} {\bibinfo {author} {\bibfnamefont {H.~P.}\ \bibnamefont
  {Breuer}}\ and\ \bibinfo {author} {\bibfnamefont {F.}~\bibnamefont
  {Petruccione}},\ }\href@noop {} {\emph {\bibinfo {title} {The theory of open
  quantum systems}}}\ (\bibinfo  {publisher} {Oxford university press},\
  \bibinfo {year} {2002})\BibitemShut {NoStop}%
\bibitem [{\citenamefont {Datta}(2005)}]{datta_quantum_transport}%
  \BibitemOpen
  \bibfield  {author} {\bibinfo {author} {\bibfnamefont {Supriyo}\ \bibnamefont
  {Datta}},\ }\href {\doibase 10.1017/CBO9781139164313} {\emph {\bibinfo
  {title} {Quantum Transport: Atom to Transistor}}}\ (\bibinfo  {publisher}
  {Cambridge University Press},\ \bibinfo {year} {2005})\BibitemShut {NoStop}%
\bibitem [{\citenamefont {Hofer}\ \emph {et~al.}(2017)\citenamefont {Hofer},
  \citenamefont {Perarnau-Llobet}, \citenamefont {Miranda}, \citenamefont
  {Haack}, \citenamefont {Silva}, \citenamefont {Brask},\ and\ \citenamefont
  {Brunner}}]{global_issue}%
  \BibitemOpen
  \bibfield  {author} {\bibinfo {author} {\bibfnamefont {Patrick~P}\
  \bibnamefont {Hofer}}, \bibinfo {author} {\bibfnamefont {Martí}\
  \bibnamefont {Perarnau-Llobet}}, \bibinfo {author} {\bibfnamefont
  {L~David~M}\ \bibnamefont {Miranda}}, \bibinfo {author} {\bibfnamefont
  {Géraldine}\ \bibnamefont {Haack}}, \bibinfo {author} {\bibfnamefont
  {Ralph}\ \bibnamefont {Silva}}, \bibinfo {author} {\bibfnamefont
  {Jonatan~Bohr}\ \bibnamefont {Brask}}, \ and\ \bibinfo {author}
  {\bibfnamefont {Nicolas}\ \bibnamefont {Brunner}},\ }\bibfield  {title}
  {\enquote {\bibinfo {title} {Markovian master equations for quantum thermal
  machines: local versus global approach},}\ }\href {\doibase
  10.1088/1367-2630/aa964f} {\bibfield  {journal} {\bibinfo  {journal} {New
  Journal of Physics}\ }\textbf {\bibinfo {volume} {19}},\ \bibinfo {pages}
  {123037} (\bibinfo {year} {2017})}\BibitemShut {NoStop}%
\bibitem [{\citenamefont {Landi}\ \emph {et~al.}(2022)\citenamefont {Landi},
  \citenamefont {Poletti},\ and\ \citenamefont
  {Schaller}}]{Review_Modern_Physics}%
  \BibitemOpen
  \bibfield  {author} {\bibinfo {author} {\bibfnamefont {Gabriel~T.}\
  \bibnamefont {Landi}}, \bibinfo {author} {\bibfnamefont {Dario}\ \bibnamefont
  {Poletti}}, \ and\ \bibinfo {author} {\bibfnamefont {Gernot}\ \bibnamefont
  {Schaller}},\ }\bibfield  {title} {\enquote {\bibinfo {title} {Nonequilibrium
  boundary-driven quantum systems: Models, methods, and properties},}\ }\href
  {\doibase 10.1103/RevModPhys.94.045006} {\bibfield  {journal} {\bibinfo
  {journal} {Rev. Mod. Phys.}\ }\textbf {\bibinfo {volume} {94}},\ \bibinfo
  {pages} {045006} (\bibinfo {year} {2022})}\BibitemShut {NoStop}%
\bibitem [{\citenamefont {Pekola}\ and\ \citenamefont
  {Karimi}(2021)}]{EFF_temp_RevModPhys.93.041001}%
  \BibitemOpen
  \bibfield  {author} {\bibinfo {author} {\bibfnamefont {Jukka~P.}\
  \bibnamefont {Pekola}}\ and\ \bibinfo {author} {\bibfnamefont {Bayan}\
  \bibnamefont {Karimi}},\ }\bibfield  {title} {\enquote {\bibinfo {title}
  {Colloquium: Quantum heat transport in condensed matter systems},}\ }\href
  {\doibase 10.1103/RevModPhys.93.041001} {\bibfield  {journal} {\bibinfo
  {journal} {Rev. Mod. Phys.}\ }\textbf {\bibinfo {volume} {93}},\ \bibinfo
  {pages} {041001} (\bibinfo {year} {2021})}\BibitemShut {NoStop}%
\bibitem [{\citenamefont {Turkeshi}\ and\ \citenamefont
  {Schir\'o}(2021)}]{Eff_temp_PhysRevB.104.144301}%
  \BibitemOpen
  \bibfield  {author} {\bibinfo {author} {\bibfnamefont {Xhek}\ \bibnamefont
  {Turkeshi}}\ and\ \bibinfo {author} {\bibfnamefont {Marco}\ \bibnamefont
  {Schir\'o}},\ }\bibfield  {title} {\enquote {\bibinfo {title} {Diffusion and
  thermalization in a boundary-driven dephasing model},}\ }\href {\doibase
  10.1103/PhysRevB.104.144301} {\bibfield  {journal} {\bibinfo  {journal}
  {Phys. Rev. B}\ }\textbf {\bibinfo {volume} {104}},\ \bibinfo {pages}
  {144301} (\bibinfo {year} {2021})}\BibitemShut {NoStop}%
\bibitem [{\citenamefont {Hill}(1994)}]{hill1994thermodynamics}%
  \BibitemOpen
  \bibfield  {author} {\bibinfo {author} {\bibfnamefont {T.L.}\ \bibnamefont
  {Hill}},\ }\href {https://books.google.co.in/books?id=Xa-yAAAAQBAJ} {\emph
  {\bibinfo {title} {Thermodynamics of Small Systems}}},\ Dover Books on
  Chemistry\ (\bibinfo  {publisher} {Dover Publications},\ \bibinfo {year}
  {1994})\BibitemShut {NoStop}%
\bibitem [{\citenamefont {Upadhyay}\ \emph {et~al.}(2021)\citenamefont
  {Upadhyay}, \citenamefont {Naseem}, \citenamefont {Marathe},\ and\
  \citenamefont {M\"ustecapl\ifmmode \imath \else \i
  \fi{}o\ifmmode~\breve{g}\else \u{g}\fi{}lu}}]{PhysRevE.104.054137}%
  \BibitemOpen
  \bibfield  {author} {\bibinfo {author} {\bibfnamefont {Vipul}\ \bibnamefont
  {Upadhyay}}, \bibinfo {author} {\bibfnamefont {M.~Tahir}\ \bibnamefont
  {Naseem}}, \bibinfo {author} {\bibfnamefont {Rahul}\ \bibnamefont {Marathe}},
  \ and\ \bibinfo {author} {\bibfnamefont {\"Ozg\"ur~E.}\ \bibnamefont
  {M\"ustecapl\ifmmode \imath \else \i \fi{}o\ifmmode~\breve{g}\else
  \u{g}\fi{}lu}},\ }\bibfield  {title} {\enquote {\bibinfo {title} {Heat
  rectification by two qubits coupled with dzyaloshinskii-moriya
  interaction},}\ }\href {\doibase 10.1103/PhysRevE.104.054137} {\bibfield
  {journal} {\bibinfo  {journal} {Phys. Rev. E}\ }\textbf {\bibinfo {volume}
  {104}},\ \bibinfo {pages} {054137} (\bibinfo {year} {2021})}\BibitemShut
  {NoStop}%
\bibitem [{\citenamefont {Palafox}\ \emph {et~al.}(2022)\citenamefont
  {Palafox}, \citenamefont {Rom\'an-Ancheyta}, \citenamefont
  {\ifmmode~\mbox{\c{C}}\else \c{C}\fi{}akmak},\ and\ \citenamefont
  {M\"ustecapl\ifmmode \imath \else \i \fi{}o\ifmmode~\breve{g}\else
  \u{g}\fi{}lu}}]{Ozur_diff_stat_PhysRevE.106.054114}%
  \BibitemOpen
  \bibfield  {author} {\bibinfo {author} {\bibfnamefont {Stephania}\
  \bibnamefont {Palafox}}, \bibinfo {author} {\bibfnamefont {Ricardo}\
  \bibnamefont {Rom\'an-Ancheyta}}, \bibinfo {author} {\bibfnamefont
  {Bar\ifmmode \imath \else \i \fi{}\ifmmode \mbox{\c{s}}\else~\c{s}\fi{}}\
  \bibnamefont {\ifmmode~\mbox{\c{C}}\else \c{C}\fi{}akmak}}, \ and\ \bibinfo
  {author} {\bibfnamefont {\"Ozg\"ur~E.}\ \bibnamefont {M\"ustecapl\ifmmode
  \imath \else \i \fi{}o\ifmmode~\breve{g}\else \u{g}\fi{}lu}},\ }\bibfield
  {title} {\enquote {\bibinfo {title} {Heat transport and rectification via
  quantum statistical and coherence asymmetries},}\ }\href {\doibase
  10.1103/PhysRevE.106.054114} {\bibfield  {journal} {\bibinfo  {journal}
  {Phys. Rev. E}\ }\textbf {\bibinfo {volume} {106}},\ \bibinfo {pages}
  {054114} (\bibinfo {year} {2022})}\BibitemShut {NoStop}%
\bibitem [{\citenamefont {Wu}\ and\ \citenamefont
  {Segal}(2009)}]{Diode:Type2_req_Dvira}%
  \BibitemOpen
  \bibfield  {author} {\bibinfo {author} {\bibfnamefont {Lian-Ao}\ \bibnamefont
  {Wu}}\ and\ \bibinfo {author} {\bibfnamefont {Dvira}\ \bibnamefont {Segal}},\
  }\bibfield  {title} {\enquote {\bibinfo {title} {Sufficient conditions for
  thermal rectification in hybrid quantum structures},}\ }\href {\doibase
  10.1103/PhysRevLett.102.095503} {\bibfield  {journal} {\bibinfo  {journal}
  {Phys. Rev. Lett.}\ }\textbf {\bibinfo {volume} {102}},\ \bibinfo {pages}
  {095503} (\bibinfo {year} {2009})}\BibitemShut {NoStop}%
\end{thebibliography}%
	

\end{document}